\documentclass{SciPost}

\hypersetup{
    colorlinks,
    linkcolor={red!50!black},
    citecolor={blue!50!black},
    urlcolor={blue!80!black}
}

\usepackage[bitstream-charter]{mathdesign}
\DeclareSymbolFont{usualmathcal}{OMS}{cmsy}{m}{n}
\DeclareSymbolFontAlphabet{\mathcal}{usualmathcal}

\fancypagestyle{SPstyle}{
\fancyhf{}
\lhead{\colorbox{scipostblue}{\bf \color{white} ~SciPost Physics }}
\rhead{{\bf \color{scipostdeepblue} ~Submission }}

\fancyfoot[C]{\textbf{\thepage}}
}

\begin{document}

\pagestyle{SPstyle}

\begin{center}{\Large \textbf{\color{scipostdeepblue}{
Extending fusion rules with finite subgroups: A general construction of $Z_{N}$ extended  conformal field theories and their orbifoldings
}}}\end{center}

\begin{center}\textbf{
Yoshiki Fukusumi\textsuperscript{1$\star$} and
Shinichiro Yahagi\textsuperscript{2} 
}\end{center}

\begin{center}
{\bf 1} Physics Division, National Center of Theoretical Sciences, National Taiwan University, Taipei 106319, Taiwan
\\
{\bf 2} Department of Physics, Faculty of Science,
The University of Tokyo, Bunkyo-Ku, Tokyo 113-0033, Japan
\\[\baselineskip]
$\star$ \href{y1224.fukusumi@gmail.com}{\small y1224.fukusumi@gmail.com}\,,\quad
$\dagger$ \href{yahagi@hep-th.phys.s.u-tokyo.ac.jp }{\small yahagi@hep-th.phys.s.u-tokyo.ac.jp }
\end{center}

\section*{\color{scipostdeepblue}{Abstract}}
\textbf{\boldmath{%
  We construct the $Z_{N}$ symmetry extended fusion ring of bulk and chiral theories and the corresponding modular partition functions with nonanomalous subgroup $Z_{n}(\subset Z_{N})$. The chiral fusion ring provides fundamental data for $Z_{N}$- graded symmetry topological field theories and also provides algebraic data for smeared boundary conformal field theories, which describe the zero modes of the extended models. For more general multicomponent or coupled systems, we also obtain a new series of extended theories. By applying the folding trick, their partition functions correspond to charged or gapped domain walls or massless renormalization group flows preserving quotient group structures. 
}}

\vspace{\baselineskip}

\noindent\textcolor{white!90!black}{%
\fbox{\parbox{0.975\linewidth}{%
\textcolor{white!40!black}{\begin{tabular}{lr}%
  \begin{minipage}{0.6\textwidth}%
    {\small Copyright attribution to authors. \newline
    This work is a submission to SciPost Physics. \newline
    License information to appear upon publication. \newline
    Publication information to appear upon publication.}
  \end{minipage} & \begin{minipage}{0.4\textwidth}
    {\small Received Date \newline Accepted Date \newline Published Date}%
  \end{minipage}
\end{tabular}}
}}
}



\vspace{10pt}
\noindent\rule{\textwidth}{1pt}
\tableofcontents
\noindent\rule{\textwidth}{1pt}
\vspace{10pt}


\section{Introduction}
\label{sec:intro}

Symmetry is one of the most fundamental aspects of contemporary physics and mathematics. Group symmetry has been studied and applied to many cases in both communities in the previous century (See \cite{Slansky:1981yr}, for example). We note two fundamental notions, chiral 't Hooft anomaly\cite{tHooft:1979rat} in high energy physics and Lieb-Shultz-Mattis theorem\cite{Lieb:1961fr} in mathematical physics, which are relevant in the successive discussions. More recently, it is famous that the notion of symmetry has been generalized by using the path integral method in \cite{Gaiotto:2014kfa}. However, we note that its quantum Hamiltonian analog treating ring or algebraic structure of the integral of motions (or nonlocal conserved quantities) and corresponding defects appeared much earlier in mathematical physical \cite{Petkova:2000ip,Graham:2003nc,Frohlich:2004ef,Frohlich:2006ch} and in condensed matter \cite{Cobanera:2009as,Cobanera:2011wn,Cobanera:2012dc}\footnote{We also note earlier studies \cite{article,Bockenhauer:1999wt} on the corresponding algebraic structures in mathematics and cite related reviews\cite{Kawahigashi:2021hds,Evans:2023nbp}.}. There exists a large body of literature, so we note an earlier review on anomaly for continuous symmetry \cite{Harvey:2005it} and recent ones on generalized symmetry\cite{McGreevy:2022oyu,Cordova:2022ruw,Bhardwaj:2023kri,Brennan:2023mmt}.

More specifically, it is widely known that the classification of conformal field theories (CFTs) has been studied based on their group structure\cite{Cappelli:1987xt}, called a simple current. In particular, the integer spin simple current which appeared in the studies on discrete torsion\cite{Vafa:1986wx,Dixon:1986jc} plays the most fundamental role in the classifications\cite{Schellekens:1989uf,Gato-Rivera:1990lxi,Schellekens:1990ys,Gato-Rivera:1991bcq,Gato-Rivera:1991bqv,Kreuzer:1993tf,Fuchs:1996rq,Fuchs:1996dd}\footnote{More precisely, the integer spin simple current extension in the literature on the classification of the corresponding modular invariant often applies the combination of the extension (addition of the $Z_{N}$ degrees of freedom to the bulk theory) and orbifolding (or quotient by $Z_{N}$ group to the extended chiral theory). To distinguish the technical term, we denote this combination of the operations as integer spin simple current orbifolding and the addition of the degrees of freedom as extension, following recent literature in theoretical physics and mathematical physics. More traditionally, this terminology is relatively common in the studies on the (monstrous) moonshine and the related vertex operator algebra. In this manuscript, we mainly focus on the addition of the $Z_{N}$ degrees of freedom and the consequent  subalgebraic structure represented as $Z_{N}$ graded symmetry topological field theory. }. Keeping in mind the correspondence between CFTs and topological quantum field theories (TQFTs)\cite{Witten:1988hf} or topological orders (TOs) in condensed matter\cite{Laughlin:1983fy}, this connection seems more reasonable. An integer spin simple current can be considered as a generalized version of the electron operator forming the wavefunctions\cite{Blok:1991zq,Read:1998ed,Schoutens:2015uia,Dorey:2016mxm,Fuji_2017,Bourgine:2024ycr}. In other words, the singlevaluedness of the wavefunction naturally results in the appearance of the integer spin simple current\footnote{More generally, simple current with half integer spin is also possible. When the underlying symmetry is the $Z_{2}$group, this corresponds to supersymmetric models. We use the term, integer spin simple current, following the literature, but both half-integer and integer cases have been treated in this manuscript}. In the terminology of string theory, the resultant TQFTs should be considered as a \emph{fractional supersymmetric} model\footnote{There exists some abuse of terminology in the term (fractional) supersymmetry (For the readers interested in this, see corresponding discussions in \cite{Fukusumi:2023psx}). In this manuscript, we identify a fractional supersymmetric model as an anomaly-free $Z_{N}$ symmetric theory or a system with the correspondence between sectors characterized by symmetry charge. We mainly follow the relaxed terminology initiated from the established literature \cite{Gliozzi:1976qd} and studies on integer spin simple currents.}  .

\begin{figure}[htbp]
\begin{center}
\includegraphics[width=1.0\textwidth]{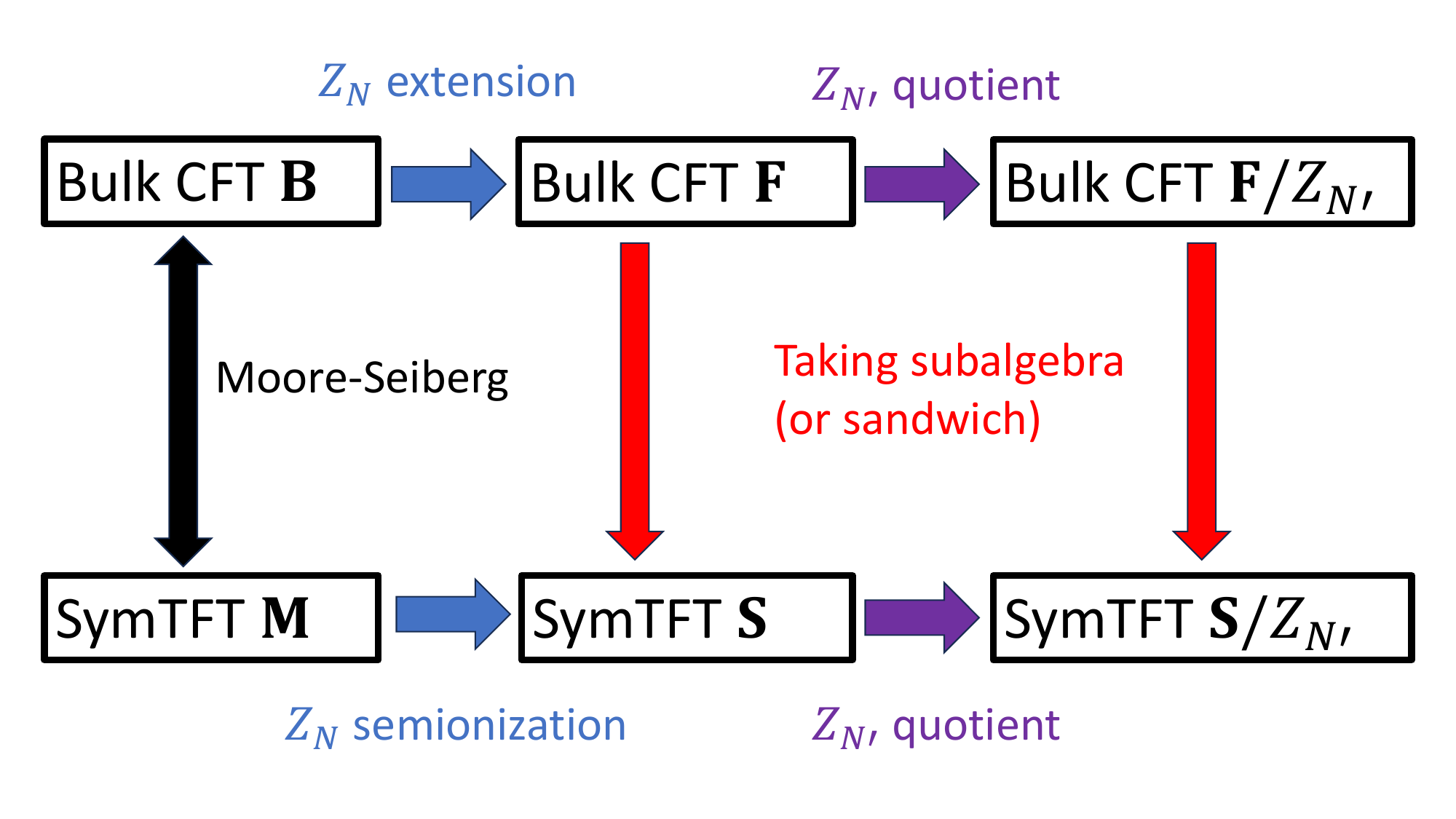}
\caption{The summary of the operations in this manuscript. Starting from a $Z_{N}$ symmetric chiral CFT or the corresponding symmetry topological field theory (SymTFT) $\mathbf{M}$, one can obtain the corresponding bulk CFT $\mathbf{B}$ or bulk fusion ring (black arrow). By extending this bosonic bulk CFT by chiral $Z_{N}$ operations or defects, one can obtain the $Z_{N}$ extended model (upper blue arrow). For the details of the chiral analog, $Z_{N}$ semionization, see the discussion in \cite{Fukusumi:2022xxe,Fukusumi_2022_c,Fukusumi:2023psx} (lower blue arrow). It should be stressed that this operation is not a straightforward $Z_{N}$ extension of the bosonic chiral algebra or MTC in the categorical formalism in \cite{turaev2000homotopyfieldtheorydimension,kirillov2001modularcategoriesorbifoldmodels,Frohlich:2003hm,mueger2004galoisextensionsbraidedtensor,etingof2009weaklygrouptheoreticalsolvablefusion}. By taking a subalgebra (or subring), one can obtain the SymTFT $\mathbf{S}$ (red arrows). By applying quotient or gauging operations to $\mathbf{F}$ or $\mathbf{S}$ by a subgroup $Z_{N'}\subset Z_{N}$, one will obtain the gauged model $\mathbf{F}/Z_{N'}$ or $\mathbf{S}/Z_{N'}$ respectively (purple arrows). The integer spin simple current orbifolding can be realized as the combinations of the lower two arrows under the condition $N=N'$. Roughly speaking, the extended model corresponds to a symmetry-enriched TO in condensed matter, and the quotient model corresponds to a quantum spin liquid in condensed matter.}
\label{extension_quotient}
\end{center}
\end{figure}

Hence, one may naively think that a part of fundamental studies on symmetry has already been established. One can see a lot of recent literature performing ``gauging" or orbifolding operation to various quantum field theories and lattice models\cite{Hamidi:1986vh,Dijkgraaf:1989hb,Vafa:1989ih,Tachikawa:2017gyf,Bhardwaj:2017xup}. However, if one looks at older literature on chiral CFT, such as Majorana CFT in the established review\cite{Ginsparg:1988ui}, or that on the corresponding structure in Moore-Read quantum Hall states ($\sim$Majorana fermion $\times U(1)$) \cite{Moore:1991ks}, one will notice a puzzle. The underlying chiral CFT and resultant partition functions are not those of a bosonic CFT\cite{Cappelli:1996np,Milovanovic:1996nj,Cappelli:2010jv}\footnote{More precisely, the characters of fractional quantum Hall effects are mock modular forms\cite{Ino:1998by,Fukusumi:2022xxe}.}. Moreover, the Majorana chiral CFT is outside of local QFT in the context of a modular tensor (or fusion) category\cite{Fukusumi:2022xxe,Fukusumi:2023psx}. They should be described by a $Z_{2}$ extended chiral CFT when constructing the corresponding chiral CFT from the Ising chiral CFT, but the procedure for obtaining such a simple current extended CFT has not been studied widely, and the literature is very limited\cite{Petkova:1988cy,Furlan:1989ra,Runkel:2020zgg,Hsieh:2020uwb}\footnote{Historically, the difficulty in realizing chiral fermion has been studied in \cite{Nielsen:1980rz,Nielsen:1981xu}, for example. The recent literature captures algebraic aspects of the difficulty.}. It should be stressed that the group extension is an inverse operation to taking a quotient, and the resultant theory can be outside of existing canonical theories. The extension naturally introduces a chiral particle (or \emph{quark}) to the system, and this operation changes the system from a bosonic theory to a quark-hadron-like particle system\cite{Goddard:1984vk,Goddard:1984hg,Fukusumi_2022_c,Fukusumi:2024ejk}. We note several recent references studying fermionizations\cite{Kitaev:2006lla,Gaiotto:2015zta,Lan_2016} and its generalization known as modular extension\cite{Lan2016ModularEO,Cho:2022kzf}. However, it should be remarked that the earlier studies on the coset conformal field theory have developed with a close connection to the fermion (or quark) representation\cite{Goddard:1984vk,Goddard:1984hg,Goddard:1986ee,Petkova:1988cy,Petkova:1988cy} with a connection to Rogers-Ramanujan type identities\cite{Andrews:1984af,Kedem:1993ze}. This is in parallel to the studies on boson-fermion correspondence in characters in string theory \cite{Gliozzi:1976qd}, and we note related reviews in statistical mechanics\cite{Campbell_2024} and string theory\cite{Harvey:2019htf}.

In this work, we complete a systematic construction of $Z_{N}$ extended CFTs and their bulk and chiral fusion rules for a general number $N$, continuing the research direction in\cite{Fukusumi:2022xxe,Fukusumi_2022_c,Fukusumi:2023psx,Fukusumi:2024cnl,Fukusumi:2024ejk}. We provide a general strategy for obtaining their bulk fusion rules (extended nonchiral fusion rings or graded fusion categories) and chiral fusion rings (extended chiral fusion rings or graded symmetry topological field theories) corresponding to the extended modular partition functions in \cite{Schellekens:1989uf,Gato-Rivera:1990lxi,Schellekens:1990ys,Gato-Rivera:1991bcq,Gato-Rivera:1991bqv,Kreuzer:1993tf,Fuchs:1996rq,Fuchs:1996dd}. Moreover, we provide a new series of modular partition functions from coupled or tensored models. By applying the folding trick to these coupled theories with symmetry $Z_{N_{1}}\times Z_{N_{2}}$, one can obtain the data of the domain wall. Surprisingly, there exists domain wall which only preserve the quotient group structure $Z_{\text{gcd}(N_{1},N_{2})}$ whereas the new modular partition functions are constructed from $Z_{\text{lcm}(N_{1},N_{2})}$ symmetry. In other words, the structure generating the coupled model and symmetry preserved through the domain wall can be different. This kind of quotient group structure appears in the studies of quantum spin liquid in condensed matter, and our method will provide a classification scheme of such systems combined with the methods called renormalization group (RG) domain wall in \cite{Brunner:2007ur,Gaiotto:2012np,Fukusumi:2024ejk,Fukusumi:2025clr}. Because quantum \emph{spin} liquid can be considered as a consequence of orbifolding of \emph{quark-hadron} models, this phenomenon itself is reasonable\cite{Sule:2013qla,Hsieh:2014lba,Chen_2017_2,Tiwari:2017wqf}. We summarize the operations and notions in this manuscript in Figure \ref{extension_quotient}. In this manuscript, we concentrate our attention on the $Z_{N}$ symmetric theories, but similar arguments on a general group $G$ will be possible by considering the extended algebra and corresponding defects and symmetries\cite{Affleck:1998nq,Fuchs:1999zi,Birke:1999ik,Birke:1999ik,Quella:2002ct}.

The rest of the manuscript is organized as follows. Section \ref{review} is a review introducing the fundamental algebraic data necessary for the later discussions. We introduce several less familiar methods and topics in the literature to provide a quantum Hamiltonian formulation of the simple current extensions. Section \ref{extension} is the main part of this work. We provide a series of data formulating $Z_{N}$ extended fusion rules for a general positive integer $N$. One can obtain a complete set of bulk fields, symmetry topological field theories (SymTFTs), chiral fields, extended boundary states, and modular partition functions when the corresponding bosonic theories $\mathbf{B}$ are given. In section \ref{counting_puzzle}, we explain a benefit of our algebraic construction in the extended or quark-hadron-like models. We demonstrate that our algebraic formalism resolves a puzzle on the operator counting problem in the extended theories and provides a general understanding of zero modes in the systems. In Section \ref{quotient}, we apply the method to a multicomponent or coupled model with $\{ Z_{N_{i}}\}_{i}$ symmetries. We provide a general method to obtain the corresponding $Z_{\text{lcm}(\{ N_{i}\})}$ extended partition functions. One can obtain a phenomenological understanding of the domain wall preserving quotient group symmetry $Z_{\text{gcd}(\{ N_{i}\})}$. Section \ref{conclusion} is the concluding remark of this work. We comment on the possible application and generalization of our methods and models in theoretical physics.

\section{General ideas for extensions and their benefit}
\label{review}

In this section, we revisit the general formalism to extend CFTs by their chiral or antichiral fusion ring. This section is an improved version of the earlier heuristic discussions in \cite{Fukusumi_2022,Fukusumi_2022_c}. One can also see related discussions in \cite{Lin:2022dhv} from a different research motivation. First thing to note is that the $Z_{N}$ extension, or combination of Fradkin-Kadanoff transformations in the $1+1$ dimensional lattice model, is \emph{not} constructed from the $Z_{N}$ symmetry Verlinde line. However, the resultant partition functions match with those constructed from the Verlinde lines when the conformal spin $s_{J}$ of the underlying $Z_{N}$ symmetry generator $J$, called $Z_{N}$ simple current, becomes a (half) integer. The simple current $J$ is a chiral primary field that behaves like a $Z_{N}$ symmetry generator, such as the fusion rule $J\times \alpha =\alpha'$ where $\alpha$ is an arbitrary label of chiral primary fields and $\alpha'$ is some field determined by the group-like property of $J$. In the following, we sometimes drop $\times$ for the application of $J$, and use the notation $\alpha'=J \alpha$, for example. The simple current with integer spin condition is called the integer spin simple current \cite{Schellekens:1989uf,Gato-Rivera:1990lxi,Schellekens:1990ys,Gato-Rivera:1991bcq,Gato-Rivera:1991bqv,Kreuzer:1993tf,Fuchs:1996rq,Fuchs:1996dd} and plays the most fundamental role in this work. More phenomenologically, the Verlinde lines are a kind of classical operation appearing in two-dimensional statistical models, and they match with the quantum analog only when the theory is anomaly-free. This provides a modern application of quantum-classical correspondence to generalized symmetry which has been first realized in \cite{Yao:2020dqx} clearly \footnote{Similar phenomenology appears commonly in literature, for example in \cite{Ryu_2012,Sule:2013qla,Chen_2017_2}, but the work \cite{Yao:2020dqx} clarified the role of discrete symmetry and anomaly evidently by performing gauging anomalous symmetry by revisiting established Fradkin-Kadanoff transformation\cite{Fradkin:1980th}. One can perform the gauging operation, but the resultant partition functions are completely outside of existing two-dimensional statistical models with Verlinde lines. This comes from the nonuniqueness of the choice of Fradkin-Kadanoff transformations, and this aspect itself has been noticed in \cite{Mong_2014,Mong:2014ova}, for example. For more recent literature, see \cite{Fukusumi_2022_c,Kawabata:2024hzx}.}. Before moving into the detailed discussions, we note a recent review\cite{Northe:2024tnm} and established reviews\cite{Fuchs:1997af,Schweigert:2000ix} for the readers unfamiliar with arguments on defects and simple currents in CFTs.

For simplicity, we consider a diagonal theory with the following partition function,
\begin{equation}
Z_{\mathbf{B}}=\sum_{\alpha} \chi_{\alpha} (\tau) \overline{\chi_{\alpha}} (\overline{\tau}),
\end{equation}
where $\chi$ ($\overline{\chi}$) is the chiral (antichiral) character and $\alpha$ is the label of primary states, and $\tau$ is the modular parameter. For later use, we introduce the modular $T$ and modular $S$ transformations as follows,
\begin{align}
&T: \tau\rightarrow \tau +1,\\
&S: \tau\rightarrow \frac{-1}{\tau}.
\end{align}
The modular $T$ transformation determines the conformal spin structure or locality of the systems, and the modular $S$ transformation determines the high-low temperature duality or stability of the theory. In other words, starting from an extended model, one can interpret the modular $S$ invariant partition function as a consequence of anyon condensations\cite{Bais:2008ni,2013PhRvX...3b1009L,Kong:2013aya}.

We also assume the ring isomorphism between chiral fusion ring $\{ \theta_{\alpha} \}$ and nonchiral fusion ring $\{ \Phi_{\alpha} \}$\cite{Fuchs:1993et,Rida:1999xu,Rida:1999ru}. This ring isomorphism is known as Moore-Seiberg data\cite{Moore:1989vd} in high energy physics or boson condensation\cite{Bais:2008ni} in mathematics. The chiral fusion ring corresponds to the modular tensor category (MTC) and the nonchiral fusion ring corresponds to the spherical fusion category with a close connection to the conformal bootstrap technique\cite{Polyakov:1974gs}\footnote{More precisely, the commutative or associative structures are not necessary to define fusion rule\cite{Nivesvivat:2025odb}. We also note related works on the vertex-operator-algebra approach to bulk CFTs\cite{Huang:2005gz,Kong:2006wa,Huang:2006ar,Moriwaki_2023,Vicedo:2025vql}.}. In the successive discussion, we use the term ``nonchiral" in a different way, so we call the nonchiral fusion ring a bulk fusion ring. 

Corresponding to the partition function $Z_{\mathbf{B}}$, we introduce the following chiral symmetry operator,
\begin{equation}
\mathcal{Q}_{\alpha}=\sum_{\beta, m} \frac{S_{\alpha,\beta}}{S_{\alpha,0}}P_{\beta, m}\otimes \overline{1}
\end{equation}
where $P$ is the projection and $m$ is the label of descendant fields and $\overline{1}$ is identity which only acts trivially on antichiral sectors. Because this is represented by the summation of projection, this operator is topological because of the commutativity with the Hamiltonian\cite{Petkova:2000ip}. More generally, one can introduce intertwining operators and the corresponding extended symmetry operators (satisfying the extended algebra\cite{Affleck:1998nq,Fuchs:1999zi,Birke:1999ik,Birke:1999ik,Quella:2002ct}) which cannot be labelled by $\{ \alpha \}$, but we restrict our attention to the objects labelled by $\{ \alpha\}$ for simplicity.

By mimicking the calculation in \cite{Petkova:2000ip} and using the Verlinde formula\cite{Verlinde:1988sn}, one can obtain the following relation,
\begin{equation}
\text{Tr} \left(\mathcal{Q}_{\alpha} e^{2i \tau (L_{0}-c/24)+2i\overline{\tau}(\overline{L_{0}}-\overline{c}/24)}\right) = \sum_{\gamma, \alpha'} N^{\gamma}_{\alpha, \beta} \chi_{\gamma} (-1/\tau) \overline{\chi_{\beta}}(-1/\overline{\tau})
\end{equation}
where $N$ is the fusion matrix with $\theta_{\alpha} \times \theta_{\beta}=N^{\gamma}_{\alpha, \beta}\theta_{\gamma}$, $L_{0}$ ($\overline{L}_{0}$) is the chiral (antichiral) Virasoro generator corresponding to the chiral (antichiral) part of the CFT Hamiltonian, and $c$ ($\overline{c}$) is the chiral (antichiral) central charge. In other words, the application of the symmetry operator $\mathcal{Q}_{\alpha}$ combined with the modular $S$ transformation $\tau \rightarrow -1/\tau$ naturally introduces the addition of a chiral particle $\theta_{\alpha}$ when considering its partition function. Hence we define the defect $\mathcal{D}_{\alpha}$ as follows,
\begin{equation}
\mathcal{D}_{\alpha}= \mathcal{S} (\mathcal{Q}_{\alpha})
\end{equation}
where $\mathcal{S}$ is the modular $S$ transformation applied to the symmetry operators.

However, we stress that the defect $\mathcal{D}_{\alpha}$, which is constructed from the modular $S$ transformation of the symmetry operator, should be distinguished from the naive application of chiral fields. It should be treated as an isolated degree of freedom, such as spin or parity structure. Only when considering the corresponding partition function does the multiplication of chiral fields appear\cite{Yao:2019bub,Fukusumi:2020irh,Runkel:2020zgg,Hsieh:2020uwb}. These additional degrees of freedom can be understood as stacking of symmetry-protected topological phases \cite{Kapustin:2017jrc,Karch:2019lnn}.

The above symmetry operators and defects satisfy the following remarkable algebraic relation, called the fusion ring,
\begin{align}
\mathcal{Q}_{\alpha} \times \mathcal{Q}_{\beta} &=\sum_{\gamma} N^{\gamma}_{\alpha,\beta} \mathcal{Q}_{\gamma}, \\
\mathcal{D}_{\alpha} \times \mathcal{D}_{\beta} &=\sum_{\gamma} N^{\gamma}_{\alpha,\beta} \mathcal{D}_{\gamma}.
\end{align}
Because these objects reflect the algebraic structure of anyons in CFTs or TQFTs, one can interpret them as the hierarchical structure of anyons. In more recent literature, this kind of symmetry algebra or the resultant category theory is called symmetry topological field theory (SymTFT)\cite{Apruzzi:2021nmk}. 

By swapping the role of the chiral and antichiral sectors, one can obtain the  corresponding antichiral symmetry operators and defects,
\begin{align}
\overline{\mathcal{Q}_{\alpha}}&=\sum_{\beta, \overline{m}} 1\otimes \frac{\overline{S}_{\alpha,\beta}}{S_{\alpha,0}}\overline{P}_{\beta, \overline{m}}, \\
\overline{\mathcal{D}_{\alpha}}&= \mathcal{S} (\overline{\mathcal{Q}_{\alpha}})
\end{align}
They also satisfy the same fusion ring relations.

Analogous to these chiral and antichiral operations, one can obtain the corresponding Verlinde line $\mathcal{Q}^{\text{V}}$\cite{Petkova:2000ip} as follows,
\begin{equation}
\mathcal{Q}^{\text{V}}_{\alpha}=\sum_{\beta, m, \overline{m}} \frac{S_{\alpha,\beta}}{S_{\alpha,0}}P_{\beta, m}\otimes \overline{P}_{\beta, \overline{m}}
\end{equation}
Hence, one can obtain the corresponding defects,
\begin{equation}
\mathcal{D}_{\alpha}^{\text{V}}= \mathcal{S} (\mathcal{Q}_{\alpha}^{\text{V}})
\end{equation}

The three objects, $\mathcal{D}$, $\overline{\mathcal{D}}$ and $\mathcal{D}^{\text{V}}$ satisfy the same fusion ring isomorphic to $\{ \theta_{\alpha}\}$. Hence, one can extend the bulk fusion ring by applying the defects in a uniform way $\{\theta_{\alpha}\} \otimes \{ \Phi_{\alpha}\}$. We stress that at this stage, we used the tensor product, not the Deligne product. Because of the ring isomorphism, one may expect the equivalence of the theory extended by these three objects. However, their actions by symmetry operators are different, and their charge conditions are different, respectively. Hence, whereas the algebraic data are the same for these extensions, their modular data can be different\footnote{In category theory, there exist similar structures, $G$-crossed extension\cite{turaev2000homotopyfieldtheorydimension,kirillov2001modularcategoriesorbifoldmodels,mueger2004galoisextensionsbraidedtensor,etingof2009weaklygrouptheoreticalsolvablefusion,Barkeshli:2014cna,davydov2021braidedpicardgroupsgraded, Bischoff:2019jho,Carqueville:2025kqs} or ribbon category\cite {Frohlich:2003hm,Heinrich:2025wkx}. However, to our knowledge, the subtelity coming from the chirality or orientation has not been discussed widely \cite{Chang:2018iay,Komargodski:2020mxz,Lin:2022dhv}.}.  

As a simplest example, let us study the $SU(2)_{1}$ Wess-Zumino-Witten (WZW) model\cite{Wess:1971yu,Witten:1983tw,Witten:1983ar}. In this theory there exist two primary fields, $\{ I, j \}$ with conformal dimension $\{ h_{I}=0, $ $ h_{j}=1/4\}$. The operator $I$ is the identity operator and $j$ is the $Z_{2}$ simple current with $j\times j=I$. The $Z_{2}$ symmetry in this theory is anomalous because of the conformal dimension $1/4$. Corresponding to this anomalous conformal dimension, the modular partition function classified by the chiral $Z_{2}$ operation $\mathcal{Q}_{j}$ and its antichiral analog $\overline{\mathcal{Q}_{j}}$ takes a different form. For the chiral extension, the extended modular partition functions are,
\begin{align}
Z_{Q=0}&=\chi_{I}(\overline{\chi_{I}}+\overline{\chi_{j}}),\\
Z_{Q=1/2}&=\chi_{j}(\overline{\chi_{I}}+\overline{\chi_{j}}),
\end{align}
where the charge $Q$ is defined by the chiral symmetry $\mathcal{Q}_{j}$. The corresponding antichiral modular partition functions classified by the antichiral symmetry operator $\overline{\mathcal{Q}_{j}}$ are,
\begin{align}
Z_{\overline{Q}=0}&=(\chi_{I}+\chi_{j})\overline{\chi_{I}},\\
Z_{\overline{Q}=1/2}&=(\chi_{I}+\chi_{j})\overline{\chi_{j}},,
\end{align}
These partition functions are not modular invariant but modular covariant. We remark that the existing partition function of a two-dimensional statistical mechanical model equivalent to $SU(2)_{1}$ WZW model with Verlinde cannot produce these partition functions. This is a variant of the discussion in \cite{Furuya:2015coa,Numasawa:2017crf,Kikuchi:2019ytf} emphasizing the $Z_{2}$ extension and the resultant modular covariance.

This kind of breaking of quantum-classical correspondence first appeared evidently in \cite{Yao:2020dqx} and its relation to quantum anomaly and chirality has been studied in\cite{Fukusumi_2022,Fukusumi_2022_c}. In other words, an anomaly-free theory is an exceptional theory satisfying the following correspondence\footnote{In the successive discussions, we note the space dimension of lattice models and the spacetime dimension of QFTs.},
\begin{equation} 
\begin{split}
&\{\text{$D+1$-dimensional statistical mechanics}\} \\
\sim &\{\text{$D$-dimensional quantum lattice}\} \\
\sim &\{\text{$D+1$-dimensional boundaries of $D+2$-dimensional TQFT}\}
\end{split}
\end{equation}
When applying the extension or gauging. For the readers in high energy physics, we remind that the above anomaly is governed by the chiral (or antichiral) 't Hooft anomaly condition or the LSM anomaly condition. For two-dimensional $Z_{N}$ symmetric CFTs, these conditions result in the appearance of an integer spin simple current. The unusual point of the above classification is that the non-anomalous theory in the context of orbifolding (or ``gauging"), such as the parafermion model\cite{Fateev:1985mm}, can be anomalous. However, by considering the construction of the wavefunction of TQFT, it is reasonable because the chiral parafermion cannot produce a single-valued wavefunction via CFT/TQFT straightforwardly. In other words, the parafermionic model is realizable in $1+1$ dimensions but unrealizable in $2+1$ dimensions, contrary to the common belief\footnote{This controversy can be resolved by considering coupling to $U(1)$ gauge field to the parafermionic model, and this model is known as the Read-Rezayi state\cite{Read:1998ed}. However, it is known that the resultant partition function breaks Moore-Seiberg data, because of the complicated mixing of flux and parafermionic particles\cite{Cappelli:1996np}.}.

\section{Extending two-dimensional conformal field theories}
\label{extension}

In this section, we demonstrate a general method extending a CFT $\mathbf{B}$ with a $Z_{N}$ symmetry where $N$ is a general positive integer. For an $N$ prime number, the previous works \cite{Fukusumi:2024cnl,Fukusumi:2024ejk} by one of the authors provide the construction, and the present method is a natural generalization of the method in \cite{Fukusumi:2024cnl}. The most fundamental structure is an anomaly-free subgroup\cite{Bhardwaj:2017xup,Tachikawa:2017gyf} and the appearance of the corresponding zero modes. It should be stressed again that whereas the algebraic construction is rigorous and evident, the resultant algebraic structure is outside of the existing modular fusion category. They will correspond to the premodular or (fractional)super fusion category in literature, but the studies on these categorical structures are still in development. Hence, we believe the algebraic data in this manuscript will be useful for the future construction of such category theories. 

For simplicity, we begin the discussion from a diagonal $Z_{N}$ symmetric model $\mathbf{B}$ where the algebraic structure of the fusion category, and the modules are governed by the same fusion ring. This condition is nothing but the Moore-Seiberg data\cite{Moore:1989vd}, and the systems are local at the begining. For simplicity, let us assume that $J$ be the $Z_{N}$ simple current such that
\begin{equation}
    \text{($N$ integer) } h_{J^k} \text{ is integer for all } k
\end{equation}
or
\begin{equation}
    \text{($N$ even) }h_{J^k} \text{ is integer for even } k \text{ and half-integer for odd } k \,.
\end{equation}
where $h_{\alpha}$ ($\overline{h}_{\alpha}$) is the chiral (antichiral) conformal dimension of a field labelled by $\alpha$. 
These conditions appear in the studies of (half) integer spin simple current or discrete torsion in older literature\cite{Schellekens:1989uf,Gato-Rivera:1990lxi,Schellekens:1990ys,Gato-Rivera:1991bcq,Gato-Rivera:1991bqv,Kreuzer:1993tf,Fuchs:1996rq,Fuchs:1996dd}. From a modern perspective, one can interpret the simple current as $Z_{N}$ generalization of electron operator in a TQFT\cite{Lu_2010,Schoutens:2015uia} and the condition for the integer spin simple current condition is nothing but single valuedness of the wavefunctions. Whereas this condition itself is the most fundamental for the construction of the wavefunction of TQFTs or TOs in condensed matter, this has not captured sufficient attention in the fields.

We write the primary field with period $n$ as
\begin{equation}
    \theta^{(n)}_{a,p} \,,\quad p\in\{0,1,\dots,n-1\}
\end{equation}
where $a$ is a label of sectors and $p$ is the $Z_{n}$ parity defined by modulo $n$ satisfying the following  relations,
\begin{equation}
    J^{n} \times \theta_{a,p} = \theta_{a,p} \,,\quad J^k\times \theta_{a,p} =\theta_{a,p+k} \neq \theta_{a,p} \text{ for all $k<n$.}
\end{equation}
Hence, for the compatibility with $Z_{N}$ symmetry, the period $n$ must be a divisor of $N$ and we denote the corresponding set as $\{n\in \mathbf{N}: n|N\}$ where $\mathbf{N}$ is the set of nonnegative integers. We denote this set $\{ n|N\}$ in short. For the readers interested in the superfusion category, we note that the number $n$ provides the classification of the generalized version of $q-$type or $m-$type object. The sectors with $n\neq N$ result in the appearance of zero modes after the extension, and this structure breaks the locality condition, as we will discuss in Section \ref{counting_puzzle}.

The spherical fusion category or bulk fusion ring \textbf{B} of the corresponding theory satisfies the following ring isomorphism or Moore-Seiberg data 
\begin{equation}
    \Phi^{(n)}_{a,p,-p}=\theta_{a,p}^{(n)}, 
\end{equation}
for all $a,\ p, \ n$. Hence as fusion rings, the equivalence $\{\theta_{a,p}^{(n)}\}=\{\Phi^{(n)}_{a,p,-p}\}$ holds trivially.

By the simple current extension, the algebra \textbf{F} consists of
\begin{equation}
    \Phi^{(n)}_{a,p,\bar{p},q} \,,\quad p,\bar{p}\in\{0,1,\dots,n-1\} ,\, q\in\{0,1,\dots,\tfrac{N}{n}-1\} \,,
\end{equation}
which obeys
\begin{align}
    & \mathcal{D}_{J} \times \Phi^{(n)}_{a,p,\bar{p},q} =
    \begin{cases}
        \Phi^{(n)}_{a,p+1,\bar{p},q} & (p=0,1,\dots,n-2) \\
        \Phi^{(n)}_{a,0,\bar{p},q+1} & (p=n-1)
    \end{cases} \\
    & \overline{\mathcal{D}}_{\bar{J}} \times \Phi^{(n)}_{a,p,\bar{p},q} =
    \begin{cases}
        \Phi^{(n)}_{a,p,\bar{p}+1,q} & (\bar{p}=0,1,\dots,n-2) \\
        \Phi^{(n)}_{a,p,0,q+1} & (\bar{p}=n-1)
    \end{cases}
\end{align}
Because of the extension, the new label $\overline{p}$ and $q$ are introduced. For the readers interested in the actual calculation of the fusion coefficient, we note a useful relation, 
\begin{equation}
\mathcal{D}_{J} (\Phi_{\alpha} \times \Phi_{\beta})=\sum_{\gamma}N^{\gamma}_{\alpha,\beta} \mathcal{D}_{J}\Phi_{\gamma}
\end{equation}
where $\alpha, \ \beta,\ \gamma $ are the label of objects. By using this relation and the fusion coefficient of the original bosonic model determined by the Verlinde formula\cite{Verlinde:1988sn}, one can systematically construct corresponding extended fusion coefficients\footnote{As far as we know, this kind of extended fusion coefficients first appeared in the studies of fermionic string theories. (See \cite{Ginsparg:1988ui}, for example).}.

When assuming the $Z_{N}$ symmetry is anomaly-free, satisfying the (half) integer spin simple current condition, the corresponding partition function is
\begin{equation}
    Z_Q = \sum_{n\in \{n|N \}} \frac{N}{n} \sum_{a:  Q_J(\theta^{(n)}_a)=Q} \left| \sum_{p=0}^{n-1} \chi_{a,p} \right|^2
\label{universal_function}
\end{equation}
where $\chi_{a,p}$ is the chiral character corresponding to the field $\theta^{(n)}_{a,p}$ and $Q_{J}(\alpha)$ is the monodromy charge defined by $\mathcal{Q}_{J}$. In a unitary mode, the monodromy charge $Q_{J}(\alpha)$ coincides with the charge defined by the conformal dimensions, $h_{J}+h_{\alpha}-h_{J\alpha}$\cite{Gannon:2003de}. It should be remarked that at the level of character, we have avoided introducing the label $q$ to compare the results in existing literature. The above partition function perfectly produces the previous results by taking $Q=0$ (for example, see reviews\cite{Fuchs:1997af,Schweigert:2000ix}) and it is consistent with the operator counting from $\mathbf{F}=\{ \Phi\}$. Hence, we propose that the above extended bulk fusion ring or extended fusion category produces the simple current extension in general. For simplicity, we concentrated our attention on the integer spin simple current, but one can easily generalize the algebraic method to the anomalous case. The corresponding modular covariants are as follows,
\begin{align}
    Z_Q &= \sum_{n\in \{ n|N\}} N \sum_{a : Q_J(\theta^{(n)}_{a,p})=Q} \chi_{a,p}  \left(\sum_{p=0}^{n-1} \overline{\chi_{a,p}} \right) \\
 Z_{\overline{Q}} &= \sum_{n\in \{ n|N\}} N \sum_{a:  Q_{\overline{J}}(\overline{\theta}^{(n)}_{a,p})=\overline{Q}}  \left(\sum_{p=0}^{n-1} \chi_{a,p} \right) \overline{\chi_{a,p}}
\end{align}
Because of the anomaly, the modular covariant takes different forms depending on the chiral or antichiral extensions\footnote{In this work, we have written down partition functions with positive amplitudes. To obtain the closed set of the partition functions under modular transformations, it is necessary to insert the $Z_{N}$ parity operator into the trace as in Appendix. \ref{twisted_fusion}.}. As a consequence, the correspondence between the modular partition functions and SymTFT is broken. However, we note again that one can still obtain the algebraic data of SymTFT in general, as we will demonstrate in the next subsection.

\subsection{Symmetry topological field theory and Deligne product: Both anomalous and anomaly-free cases}
In this subsection, we provide algebraic data of symmetry topological field theory (SymTFT) or the extended chiral algebra of the $Z_{N}$ symmetric model. We mainly follow the approach in \cite{Fukusumi:2024cnl}, and use the notation in the previous section. We note that the integer spin simple current condition is not necessary for the algebraic discussions in this section, except for Eq. \eqref{relaxed_correspondence}\footnote{One can see related arguments in \cite{Duan:2023ykn,Chen:2023jht}.}.

By applying the anyon condensation, the algebraic data of SymTFT \textbf{S} is
\begin{equation}
    \Psi^{(n)}_{a,p,q} = \frac{1}{\sqrt{Nn}} \sum_{p'=0}^{n-1} \Phi^{(n)}_{a,p+p',-p',q} \,,\quad p\in\{0,1,\dots,n-1\} ,\, q\in\{0,1,\dots,\tfrac{N}{n}-1\} \,.
\end{equation}
They form a subring of the $Z_{N}$ extended fusion ring $\mathbf{F}$. One can see the corresponding phenomenological arguments in \cite{Huang:2023pyk,Bhardwaj:2024ydc,Huang:2024ror,KNBalasubramanian:2025vpe}. Because the objects in $\mathbf{S}$ are graded by $Z_{N}$, one can easily apply the $Z_{N}$ orbifolding procedure to $\mathbf{S}$. Hence, by restricting our attention to the modular invariant sector $Q=0$, this quotient operation implements the integer spin simple current orbifolding in the literature. 

By changing the basis, one can also obtain the following extended bosonic SymTFT,
\begin{equation}
   \Theta_{a,p,r}^{(n)}=\sum_{q=0}^{\frac{N}{n}-1} \frac{\omega^{nqr}}{\sqrt{N/n}} \Psi^{(n)}_{a,p,q} 
\end{equation}
where $\omega$ is a $N$-th root of unity, $\omega=e^{\frac{2\pi i}{N}}$. 

This provides an algebraic representation of the extended theory. For the readers unfamiliar with the extended algebra, we note that the above ring should be interpreted as a symmetry operator in $2+1$ dimensional system. In a bosonic theory, there exists the following correspondence,
\begin{equation} 
\begin{split}
&\{\text{$2$-dimensional BCFT}\} \\
\sim &\{\text{Sym TFT of $2+1$-dimensional TQFT}\} \\
\sim &\{\text{Chiral algebra of $2$-dimensional CFT}\}
\end{split}
\end{equation}
This is an alternative expression of Moore-Seiberg data\cite{Moore:1988qv,Moore:1988ss} or SymTFT/TO correspondence in the fields\cite{Chatterjee:2022kxb}\footnote{For the readers interested in these fields, we note a review\cite{Kong:2022cpy} and reference therein. Here we note several earlier works\cite{Kong:2013aya,Kong:2014qka,kong2015boundarybulkrelationtopologicalorders,Lan_2016,Lan2016ModularEO,Kong:2017etd,Kong:2017hcw}.}. In short, one can summarize the correspondence as follows,
\begin{equation}
\{\theta\}=\{ \Theta\} =\{ \Phi\}
\end{equation}
with $r=0$.

However, in an extended model, the correspondence of the anomaly-free theory should be relaxed as follows,
\begin{equation} 
\begin{split}
&\{\text{$2$-dimensional smeared BCFT}\} \\
\sim &\{\text{Sym TFT of $2+1$-dimensional TQFT}\} \\
\sim &\{\text{Chiral algebra of $2$-dimensional CFT}\}
\end{split}
\label{relaxed_correspondence}
\end{equation}
The smeared BCFT has appeared evidently in the pioneering work \cite{Cardy:2017ufe} by Cardy\footnote{More precisely, the correspondence between gapped phases and BCFTs has already been noticed in the late $1980$s in the study of integrable models\cite{Date:1987zz,Saleur:1988zx}. One can see related historical aspects in \cite{Foda:2017vog,Fukusumi:2024ejk}. Whereas BCFTs apply to the analysis for boundary critical phenomena\cite{Cardy:1986gw,Cardy:1989ir}, smeared BCFTs can be outside of boundary critical phenomena in the usual sense.}. Regardless of its applicability to a large class of gapped phases, applications of smeared BCFT are relatively less common\cite{Lencses:2018paa,Ares:2020uwy,Li:2022drc,Cordova:2022lms,Kikuchi:2021qxz,Kikuchi:2022biw,Kikuchi:2022gfi,Kikuchi:2022ipr,Fukusumi:2024ejk,Fukusumi:2025clr,Wen:2025xka,Choi:2025ebk}. The extension of smeared BCFTs applicable to the SSB phase has been proposed in \cite{Fukusumi:2024ejk} for the $N$ prime case. It is reasonable to expect that the same phenomenology is true for higher-dimensional systems by replacing chiral CFT with ancillary CFT\cite{Nishioka:2022ook} and applying the higher condensation theory in \cite{Kong:2024ykr}, but this is out of the scope of the present manuscript. Consequently, we can summarize the relaxed correspondence as follows,
\begin{equation}
\{\phi\}=\{ \Psi\} \subset \{ \Phi\}
\end{equation}
where their indices take arbitrary values and $\subset$ is the inclusion as subalgebra.

One can obtain a chiral fusion ring from the following CFT/TQFT correspondence or ring isomorphism,
\begin{equation}
\{ \Psi^{(n)}_{a,p,q} \}=\{ \phi^{(n)}_{a,p,q} \}
\end{equation}
where $\phi$ is the extended chiral fields. By applying the basis transformation, one can obtain the following expression, 
\begin{equation}
\{ \Theta^{(n)}_{a,p,r} \}=\{ \theta^{(n)}_{a,p,r} \}
\end{equation}
By choosing $r=0$, the above chiral fields $\{ \theta\}$ reproduce the bosonic chiral fusion ring or MTC. Corresponding to this representation, one can write down the corresponding smeared Cardy states as follows,
\begin{equation}
| \theta^{(n)}_{a,p,r}\rangle =\sum_{q=0}^{\frac{N}{n}-1} \frac{\omega^{nqr}}{\sqrt{N/n}} |\phi^{(n)}_{a,p,q}\rangle 
\end{equation}
where $|\phi^{(n)}_{a,p,q}\rangle$ satisfy Cardy's conditions determined by the fusion algebra of $\{ \Psi \}$ or $\{ \Theta\}$. This is an improved version of smeared BCFTs in \cite{Fukusumi:2024ejk}. We also note related earlier works\cite{Smith:2021luc,Weizmann,Fukusumi:2021zme}. The above Cardy states reproduce $Z_{N}$ invariant Cardy states realizable at the boundary of $1+1$ dimensional spin or particle systems by taking $r=0$.

One can apply the formal chiral-antichiral decomposition to the bulk fusion ring by introducing these extended chiral fields as follows,
\begin{equation}
\Phi_{a,p, \overline{p}, q}^{(n)}=\sum_{q'=0}^{\frac{N}{n}-1} \frac{\phi^{(n)}_{a,p,q'}\otimes_{\text{D}}\overline{\phi}^{(n)}_{a,\overline{p},q-q'}}{\sqrt{N/n}}
\end{equation}
where $\otimes_{\text{D}}$ is the Deligne or box product in literature\cite{Deligne1990} and it should be distinguished from the usual tensor product. Because its explicit construction for an extended model has not been achieved in our knowledge, we have used unusual notation $\otimes_{\text{D}}$. We leave this as an open problem, but we stress that we determined the algebraic and modular data of the chiral and bulk fusion ring in this work. Hence, the corresponding well-defined category theory should respect these data, and this will provide a clue in constructing a new series of (premodular or fractional supersymmetric) category theories. It should be kept in mind that abstract algebra appeared earlier than category theory, and category theory itself has been constructed with a close connection to abstract algebra.  

For the readers interested in the bulk and chiral correlation functions, we note the set of bulk fields as follows,
\begin{equation}
\Phi_{a,p, \overline{p}, q}^{(n)}(z, \overline{z}) =\sum_{q'=0}^{\frac{N}{n}-1} \frac{\phi^{(n)}_{a,p,q'}(z)\otimes_{\text{D}}\overline{\phi}^{(n)}_{a,\overline{p},q-q'}(\overline{z})}{\sqrt{N/n}}
\end{equation}
More practically, one can obtain the corresponding multipoint correlation functions by combining the relation between operator-product-expansion and the bulk fusion ring and conformal bootstrap\cite{Fuchs:1993et,Rida:1999xu,Rida:1999ru,Nivesvivat:2025odb}. By applying the doubling trick $\overline{\phi} \rightarrow \phi$ to this set of bulk fields\cite{PhysRevLett.54.1091,Cardy:1986gw}, one can obtain the set of chiral fields as follows,
\begin{equation}
\Phi_{a,p_{1}, p_{2}, q}^{(n)}(z_{1}, z_{2}) =\sum_{q'=0}^{\frac{N}{n}-1} \frac{\phi^{(n)}_{a,p_{1},q'}(z_{1})\phi^{(n)}_{a,p_{2},q-q'}(z_{2})}{\sqrt{N/n}}
\end{equation}
for $n \neq N$ and 
\begin{equation}
\phi^{(N)}_{a,p,q}(z)
\end{equation}
for $n=N$. As one can easily see from these expressions, the fields, except for the sector $n=N$, are two-point objects and nonlocal because they are constructed from disorder operators or sectors containing zero modes. This is a reasonable consequence because the theory should be outside of existing MTCs. We remind here again that the chiral quark-hadron-like system, such as a chiral Majorana fermion system, should have difficulties in realizing a lattice model \cite{Nielsen:1980rz,Nielsen:1981xu}, and this nontriviality of the theory is reasonable from this historical view. They will be described by premodular fusion categories or (factional) superfusion categories, and the resultant category should break some conditions corresponding to the axiom of local QFT. The nonlocality of the sector is analogous to entangled paired structure in quantum information theory\cite{Stern2003GeometricPA}, and the corresponding phenomena have been studied in the literature on condensed matter (for more detailed discussions on this structure, see \cite{Fukusumi:2023psx}).

\section{Operator counting: A puzzle solved by extended chiral fusion ring}
\label{counting_puzzle}

In this section, we explain a puzzle solved by our method to demonstrate the benefit of our formalism. The puzzle is a compatibility between chiral-antichiral decomposition and bulk operator counting only by assuming the bosonic chiral (or antichiral) fields $\{ \theta^{(n)}_{a,p,0}\}$ (or $\{ \overline{\theta}^{(n)}_{a,\overline{p},0}\}$). We stress that this kind of operator counting problem comes from zero modes (or a kind of disorder fields), and it is necessary to study such objects more carefully when treating them as algebraic objects\footnote{We note that the corresponding ambiguity of the definition of chiral and bulk structure has already been noticed in literature on integer spin simple current (See section $8.4$ of \cite{Schellekens:1996tg}, for example). The discussion in this section provides a more evident and general expression resolving such ambiguities.}. Phenomenologically, the arguments in this section provide an elementary understanding of zero modes or degeneracies of wavefunctions in TQFTs (or TOs)\cite{Moore:1991ks}. In more recent terminology, the resultant models correspond to symmetry enriched topological orders (SETs), and our analysis demonstrates that more generalized category theories are required for the consistent description of the zero modes.

First, let us concentrate our attention to $\{ \theta^{(n)}_{a,p,0}\}_{p=0}^{n-1}$ fixing the number $i$ and $n(\neq N)$ in a $Z_{N}$ anomaly free theory. In the bosonic theory, there exists $n$ bulk fields, $\{ \Phi_{a,p,-p}^{(n)} \}_{p=0}^{n-1}$ and they correspond to sectors in bosonic modular invariant $Z_{\mathbf{B}}$. Hence one can consider the condensation from the $n^{2}$ objects $\{\theta^{(n)}_{a,p,0}\}_{p} \otimes \{ \overline{\theta}^{(n)}_{a,\overline{p},0}\}_{\overline{p}} $ to $\{ \Phi_{a,p,-p}^{(n)} \}_{p}$ and this is the usual boson condensation\cite{Bais:2008ni,Kong:2013aya}. Because of the condensation, the tensor product $\otimes$ becomes the Deligne product $\otimes_{\text{D}}$, and this has already been established in existing literature. Schematically, one can write down as follows,
\begin{equation}
\{\theta^{(n)}_{a,p,0}\}_{p} \otimes \{ \overline{\theta}^{(n)}_{a,\overline{p},0}\}_{\overline{p}} \Rightarrow \{ \theta^{(n)}_{a,p,0}\otimes_{\text{D}} \overline{\theta}^{(n)}_{a,-p,0}\}_{p}=\{ \Phi_{a,p,-p}^{(n)} \}_{p}
\end{equation}
where $\Rightarrow$ represents the boson condensation\footnote{It should be noted that the algebraic data of $\{ \Phi\}$ is determined only from the single copy of $\{ \theta\}$ by Moore-Seiberg data. In other words, the Deligne product (or corresponding theory describing a condensation) should be a natural generalization of this ring isomorphism $\{ \Phi\}=\{ \theta\}$.}. This is a reduction from $n^{2}$ objects to $n$ objects.

However, when applying the $Z_{N}$ extension to this theory, the corresponding sector of bulk fields labelled by $a$ contains $nN$ objects. Hence, the naive $\{\theta^{(n)}_{a,p,0}\}_{p} \otimes \{ \overline{\theta}^{(n)}_{a,\overline{p},0}\}_{\overline{p}} $ decomposition constrcuted from the bosonic theory fails in this case, because the number $n^{2}$ is smaller than $nN$. To explain this extended bulk theory, it is necessary to study the condensation from $\{\theta^{(n)}_{a,p,r}\}_{p,r} \otimes \{ \overline{\theta}^{(n)}_{a,\overline{p},\overline{r}}\}_{\overline{p}, \overline{r}} $ or equivalent   $\{\phi^{(n)}_{a,p,q}\}_{p,q} \otimes \{ \overline{\phi}^{(n)}_{a,\overline{p},\overline{q}}\}_{\overline{p}, \overline{q}} $ theory. The operator counting of these theories is $N^{2}$, and this number is greater than $nN$, that of the corresponding bulk fields. Hence, our discussion successfully explains the operator counting of bosonic and extended theories, and the phenomenology is schematically summarized as follows,
\begin{equation}
\{\theta^{(n)}_{a,p,r}\}_{p,r} \otimes \{ \overline{\theta}^{(n)}_{a,\overline{p}, \overline{r}}\}_{\overline{p},\overline{r}}=\{\phi^{(n)}_{a,p,q}\}_{p,q} \otimes \{ \overline{\phi}^{(n)}_{a,\overline{p},\overline{q}}\}_{\overline{p}, \overline{q}} \Rightarrow \{ \Phi_{a,p,\overline{p}, q}^{(n)} \}_{p,\overline{p},q}
\end{equation}
where $\Rightarrow$ represents anyon condensation or introduction of the conjectural Deligne product decomposition of the bulk fields. This is a reduction from $N^{2}$ objects to $nN$ objects and is clearly different from the usual boson condensation. The corresponding counting problem can be seen in \cite{Lou:2020gfq} and other works by one of the authors\cite{Fukusumi_2022_c,Fukusumi:2022xxe,Fukusumi:2023psx,Fukusumi:2024cnl}.

For the readers interested in the consequence of this counting, we remind them that the zero modes can appear as degeneracies in a massive renormalization group flow. Based on the phenomenology of smeared BCFT\cite{Cardy:2017ufe} determined by the algebraic structure of $\{ \phi \}$, the above analysis demonstrates the existence of nontrivial zero modes in quark-hadron-like systems in general.

\section{General method for the extension of tensored models or domain wall preserving quotient group symmetry}
\label{quotient}
\subsection{Extension by lowest common multiple: Partition function of coupled models}

When studying a single copy of a theory, the argument in the previous section provides a general class of extended conformal field theories. However, when studying two or multiple copies of CFTs, there exist more unusual extensions\cite{Fukusumi:2024ejk}. 

First, we write down the general form of the partition function of multiple copies of models with different group symmetries. Let us assume there exist CFTs with $Z_{N_{i_{+}}}$ symmetry where $i_{+}$ takes value from $1$ to $M_{+}$ and those with $Z_{N_{i_{-}}}$ symmetry where $i_{-}$ takes value from $1$ to $M_{-}$.
Then one can introduce $Z_{\text{lcm}(\{N_{i_{+}}, N_{i_{-}} \})}$ extension by $J = \prod_{i_{+}=1}^{M_{+}}j_{i_{+}}\prod_{i_{-}=1}^{M_{-}}\overline{j_{i_{-}}}$. We also assume the integer spin simple current condition for the conformal spin $s_{J}=\sum_{i_{+}=1}^{M_{+}}h_{j_{i_{+}}}-\sum_{i_{-}=1}^{M_{-}}h_{j_{i_{-}}}$. The resultant construction of the models is the same as that in the previous section. Hence, one can represent the partition function as in \eqref{universal_function}, but the nontrivial mixing of chirality will occur corresponding to the lower index $+$ and $-$ of $i$. The resultant partition function is written down in the following way,
\begin{equation}
    Z_{Q}^{\text{ex}} = \sum_{n\in \{n|N \}} \frac{N}{n} \sum_{a: Q_J(\theta^{(n)}_a)=Q} \left| \sum_{p=0}^{n-1} \Xi_{a,p} \right|^2
\label{universal_nonchiral}
\end{equation}
where $N=\text{lcm}(\{N_{i_{+}}, N_{i_{-}} \})$ is the lowest common multiple of $\{ N_{i_{+}}\}$ and $\{ N_{i_{-}} \}$ and the nonchiral character $\Xi_{a,p}$ is constructed from the product of general character $\Xi_{\{ \alpha\}} = \prod_{i_{+}=1}^{M_{+}}\chi_{\alpha_{i_{+}}}\prod_{i_{-}=1}^{M_{-}}\overline{\chi}_{\alpha_{i_{-}}}$ by applying the classification of $J$ and $\theta_{a,p}^{(n)}$ is the corresponding combination of chiral and antichiral fields\footnote{These fields $\{ \theta \}$ containing nontrivial mixing of chiral and antichiral fields are called nonchiral anyon\cite{Kong:2019cuu} and played a role in solving the inconsistency between theory and experimental observation of thermal Hall conductance\cite{Barkeshli:2015afa,Son:2015xqa,Wang2017TopologicalOF,Mross_2018}. We also noted that the corresponding phenomena were predicted earlier in \cite{Stone:2012ud} based on the established arguments on the anomaly-inflow mechanism in high-energy physics\cite{Callan:1984sa}. }. This provides a generalized version of modular invariants in \cite{Fukusumi:2024ejk}. For nontrivial coupling, the condition $\text{gcd}(\{N_{i_{+}}, N_{i_{-}} \})\neq 1$ is required as we demonstrate below. This condition restricts the forms of symmetries of the coupled models. One can also introduce the corresponding anomalous model as follows,
\begin{align}
    Z_{Q}^{\text{ex}} &= \sum_{n\in \{ n|N\}} N \sum_{a: Q_J(\theta^{(n)}_{a,p})=Q} \Xi_{a,p}  \left(\sum_{p=0}^{n-1} \overline{\Xi_{a,p}} \right) \\
 Z_{\overline{Q}}^{\text{ex}} &= \sum_{n\in \{ n|N\}} N \sum_{a: Q_{\overline{J}}(\overline{\theta}^{(n)}_{a,p})=\overline{Q}}  \left(\sum_{p=0}^{n-1} \Xi_{a,p} \right) \overline{\Xi_{a,p}}
\end{align}
One can apply the same construction of extended bulk fusion rings, SymTFT, and chiral fusion rings by considering the tensor product of the theories and their extension by $J$, both in the anomalous and anomaly-free cases, as in \cite{Fukusumi:2024ejk}.

For simplicity, let us consider two copies of anomalous $Z_{N_{1}}$ and $Z_{N_{2}}$ symmetric conformal field theories. In this setting one can consider $Z_{\text{lcm}(N_{1},N_{2})}$ symmetry generated by the $J=j_{1}j_{2}$ or $J=j_{1}\overline{j_{2}}$ where $j_{i}$ is the $Z_{N_{i}}$ simple currents respectively and $\text{lcm} (A,B)$ is the lowest common multiple of a number $A$ and $B$. For simplicity, we concentrate the discussion on the case with $J=j_{1}j_{2}$, but by replacing the chirality of one theory, one can apply the same method to the case with $J=j_{1}\overline{j_{2}}$.

From this extension, the charge defined by $J$ is extended to $Q=k/\text{lcm}(N_{1},N_{2})$ where $k$ is an integer defined by modulo $\text{lcm}(N_{1},N_{2})$. When restricting our attention to the charge zero sector which is closely related to symmetry preserving domain wall\cite{Gaiotto:2012np,Kong:2019cuu,Kaidi:2021gbs,Fukusumi:2024ejk,Fukusumi:2025clr}, the form of the charge of respective theory is restricted to $Q_{1}=\mathcal{q}/\text{gcd}(N_{1},N_{2})$ $Q_{2}=-\mathcal{q}/\text{gcd}(N_{1},N_{2})$ where $\mathcal{q}$ is an integer by modulo $\text{gcd}(N_{1},N_{2})$. Hence, to obtain a nontrivial extension compared with the original extension by $j_{i}$, it is necessary to choose $\text{gcd}(N_{1},N_{2})\neq 1$. Moreover, construction of anomaly free $Z_{\text{lcm}(N_{1},N_{2})}$ theory from anomalous $Z_{N_{1}}$ and $Z_{N_{2}}$ theories require the condition $\text{gcd}(N_{1},N_{2})\neq 1$, because of the anomaly-free condition $Q_{J}(J)$ $=Q_{j_{1}}(j_{1})+Q_{j_{2}}(j_{2})$ $=0$. This condition requires the conformal dimension of the simple currents to be $h_{j_{i}}$ $=k_{i}$ $/\text{gcd}(N_{1},N_{2})$ where $k_{i}$ is an integer or half integer satisfying the integer spin simple current condition for $J$.

It should be stressed that the condition $\text{gcd}(N_{1},N_{2})\neq 1$ require the common subgroup structure $Z_{\text{gcd}(N_{1},N_{2})}$ of the theories, but the extension we introduced is different from this subgroup. By applying the folding trick\cite{Wong:1994np} and the discussion on the RG domain wall\cite{Brunner:2007ur,Gaiotto:2012np}, one can observe Eq.\eqref{universal_nonchiral} giving the algebraic data of homomorphism $\rho$ between the theory with the $Z_{N_{1}}$ symmetry and that with the $Z_{N_{2}}$ symmetry. Moreover, by a consistency check, one can observe that \emph{the preserved symmetry under $\rho$ and the symmetry that generates the extension of the folded model, Eq.\eqref{universal_nonchiral} , are different} in this case. More precisely, the homomorphism $\rho$ preserves the \emph{common quotient group} structure $Z_{\text{gcd}(N_{1},N_{2})}$ whereas the extension has been generated from $Z_{\text{lcm}(N_{1},N_{2})}$ symmetry. We present the corresponding discussions in the next subsection.

\subsection{Homomorphism from greatest common divisor: Domain wall preserving common quotient group structure}

First, we restrict our attention to the homomorphism $\rho:$ $Z_{N_{1}}\rightarrow Z_{N_{2}}$ and assume that the following relation holds,
\begin{equation}
\rho(j_{1})=\overline{j_{2}}
\end{equation}
This comes from that the extension in Eq.\eqref{universal_nonchiral} has been performed by $J=j_{1}j_{2}$. This homomorphism comes from the arguments in \cite{Brunner:2007ur,Gaiotto:2012np}.
It is remarkable that the chirality is changed under the folding trick\cite{Wong:1994np}. Because of this condition, the ring homomorphism acts as a group homomorphism when restricting our attention to $Z_{N_{1}}$ simple current. Because of the property of group homomorphism, the following should also hold,
\begin{equation}
\rho(j_{1}^{N_{1}})=\rho (I_{1})=I_{2}, \ \rho(j_{1}^{N_{2}})=\overline{j_{2}}^{N_{2}}=I_{2}
\end{equation}
Hence, it is necessary to obtain the following for consistency,
\begin{equation}
\rho(j_{1}^{\text{gcd}(N_{(1)},N_{(2)})})=I_{2}
\end{equation}
but it contradicts with the relation $\rho(j_{1}^{\text{gcd}(N_{1},N_{2})})=\overline{j_{2}}^{\text{gcd}(N_{1},N_{2})}$ resulted from $\rho(j_{1})=\overline{j_{2}}$ except for the case with $N_{2}=\text{gcd}(N_{1},N_{2})$. However, by interpreting $j_{i}$ as object in the quotient group $\{ I, j_{i}, j_{i}^{2} ... , j_{i}^{\text{gcd}(N_{1},N_{2})-1} \} \sim Z_{\text{gcd}(N_{1},N_{2})}$, one can resolve this inconsistency preserving the relation $\rho(j_{1})=\overline{j_{2}}$ in general. More generally, by redefining the objects in the common quotient group $Z_{N'}$ where $N'$ is a common divisor of $N_{1}$ and $N_{2}$, the mapping becomes consistent. For simplicity, we concentrate our attention on the case with $N'=\text{gcd}(N_{1},N_{2})$, because this generates the largest nontrivial common quotient group. 

The structure $\{ I, j_{i}, j_{i}^{2} ... , j_{i}^{\text{gcd}(N_{1},N_{2})-1} \} \sim Z_{\text{gcd}(N_{1},N_{2})}$ or its subgroup structure is preserved under the homomorphism. We stress again that the structure $\{ I, j_{i}, j_{i}^{2} ... , j_{i}^{\text{gcd}(N_{1},N_{2})-1} \} $ is not subgroup of $Z_{N_{i}}$. This should be called common quotient group of $Z_{\text{gcd}(N_{1},N_{2})}$ $=Z_{N_{1}}/Z_{N_{1}/\text{gcd}(N_{1},N_{2})}$ $=Z_{N_{2}}/Z_{N_{2}/\text{gcd}(N_{1},N_{2})}$. Hence the partition function provides the information of $Z_{\text{gcd}(N_{1},N_{2})}$ symmetric models after the orbifolding of each theory whereas the partition function itself has been constructed from the $Z_{\text{lcm}(N_{1},N_{2})}$ extension. It seems unusual, but the appearance and usefulness of extended models in BCFTs and defect CFTs have appeared widely in the study of symmetry-breaking boundary conditions\cite{Fuchs:1999zi,Birke:1999ik,Fuchs:1999xn,Quella:2002ct,Ishikawa:2002wx,Ishikawa:2005ea} or new boundary conditions in condensed matter\cite{Affleck:1998nq,Behrend:2000us,Iino:2020ipa}. For the readers unfamiliar with the discussion, we note a textbook \cite{Recknagel:2013uja}. We stress that the partition function Eq.\eqref{universal_nonchiral} is that of the extended model (or extension of the $A$-type model), and the orbifolded model corresponds to the $D$-type model. Hence, the BCFT of our model will provide systematic data that are difficult to read off only from the $D$-type model. We leave more detailed calculations which apply to respective high-energy or condensed matter phenomena as a future problem, but we note the benefit of the simple current extension in the previous sections because the necessary algebraic data have already been obtained in the previous sections.

One of the most fundamental points in this setting is that all of the objects in $Z_{N_{1}}$ and $Z_{N_{2}}$ extended CFTs, $\mathbf{F}_{1}$ and $\mathbf{F}_{2}$, are graded by $Z_{N_{1}}$ or $Z_{N_{2}}$ by their parity. Hence, the above discussions on the quotient structure can be generalized to all of the objects in $\mathbf{F}_{1}$ and $\mathbf{F}_{2}$, and one can obtain the following mapping,
\begin{equation}
\rho: \mathbf{F}_{1}/Z_{N_{1}/\text{gcd}(N_{1},N_{2})}\rightarrow \mathbf{F}_{2}/Z_{N_{2}/\text{gcd}(N_{1},N_{2})}
\end{equation}
where $\mathbf{F}_{i}/Z_{N_{i}/\text{gcd}(N_{1},N_{2})}$ is introduced by applying the following equivalence relations,
\begin{equation}
(\mathcal{D}_{j_{i}})^{\text{gcd}(N_{1},N_{2})}\Phi_{\alpha_{i}}=(\mathcal{\overline{D}}_{\overline{j}_{i}})^{\text{gcd}(N_{1},N_{2})}\Phi_{\alpha_{i}}=\Phi_{\alpha_{i}}, 
\end{equation}
where $\alpha_{i}$ is the arbitrary label in $\mathbf{F}_{i}$. By taking a subalgebra (or sandwich construction) and applying the equivalence relation\cite{Fukusumi:2024cnl}, one can also obtain the mapping 
\begin{equation}
\rho: \mathbf{S}_{1}/Z_{N_{1}/\text{gcd}(N_{1},N_{2})}\rightarrow \mathbf{S}_{2}/Z_{N_{2}/\text{gcd}(N_{1},N_{2})}
\end{equation}
with the following equivalence relations,
\begin{equation}
(\mathcal{D}_{j_{i}})^{\text{gcd}(N_{1},N_{2})}\Psi_{\alpha_{i}}=\Psi_{\alpha_{i}}, 
\end{equation}
Hence, we conclude that the partition function Eq.\eqref{universal_nonchiral} with $Q=0$ corresponds to a gapped domain wall $\rho$ between two TQFTs (or RG domain wall between two CFTs). More generally, we propose the following phenomenology,
\begin{equation}
\text{$Z_{Q}^{\text{ex}}$ provides the data of conformal interfaces between $ \mathbf{F}_{i}/Z_{N_{i}/\text{gcd}(N_{1},N_{2})}$}
\end{equation}
or equivalently,
\begin{equation}
\text{$Z_{Q}^{\text{ex}}$ provides the data of charged domain walls between $ \mathbf{S}_{i}/Z_{N_{i}/\text{gcd}(N_{1},N_{2})}$}
\end{equation}
For more detailed data of $\rho$ and charged domain walls, one needs to detect the corresponding conformal interface by using the knowledge of coset CFTs or fusion ring homomorphisms (or Witt equivalent relation between theories). For this purpose, we note studies on conformal interfaces \cite{Quella:2006de,Kimura:2014hva,Kimura:2015nka}, on RG domain walls\cite{Stanishkov:2016pvi,Poghosyan:2022mfw,Poghosyan:2023brb} and on gapped domain walls \cite{Lan:2014uaa,Wan:2016php,Zhao:2023wtg,Jia:2025yph,Buican:2025zpm}. We also note recent related works \cite{Cordova:2025eim,Antinucci:2025uvj,Antinucci:2025fjp,KNBalasubramanian:2025vum} based on the generalized symmetry that provide a phenomenological understanding of the domain wall problems.

\begin{figure}[htbp]
\begin{center}
\includegraphics[width=1.0\textwidth]{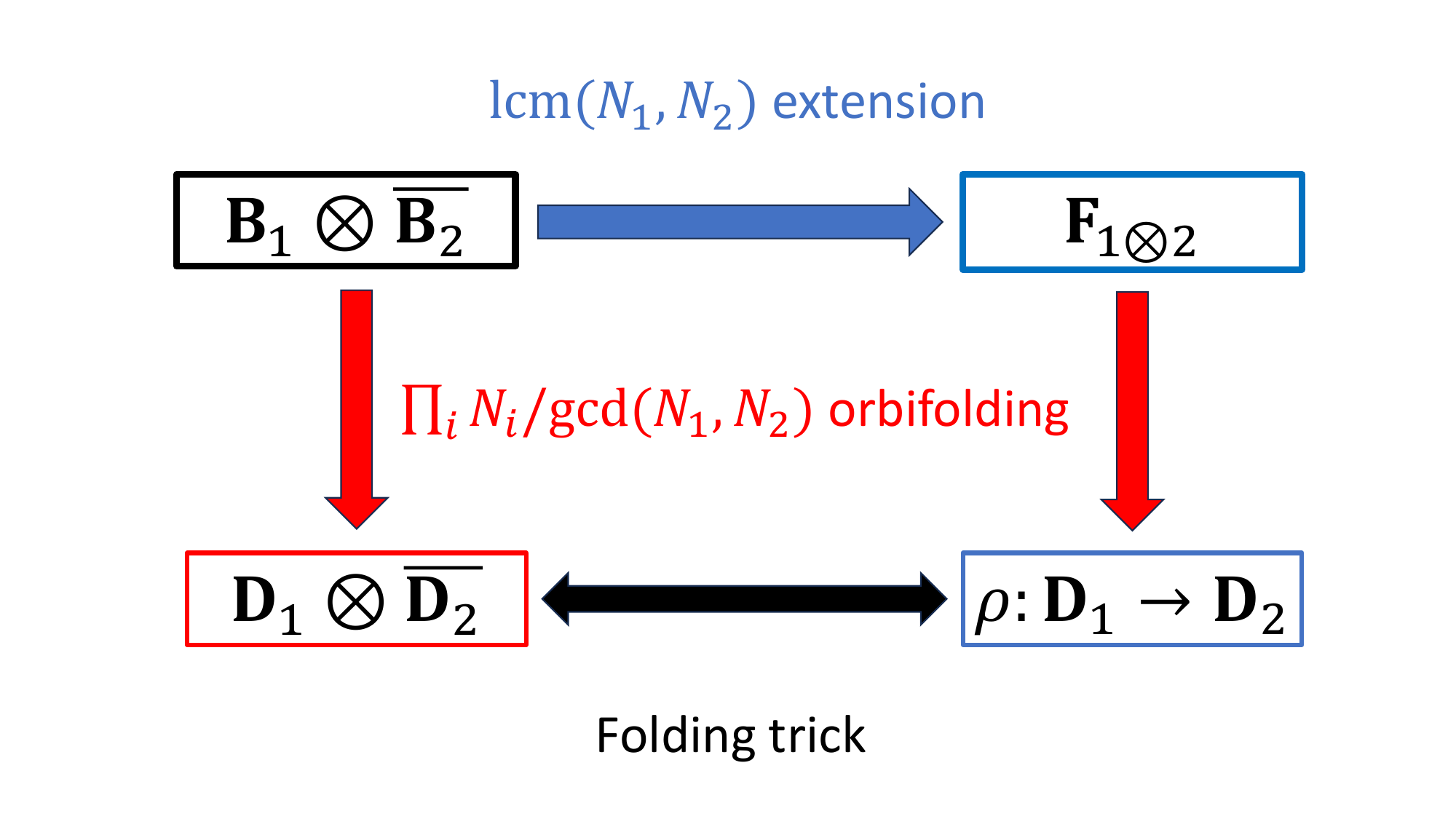}
\caption{The relationship between theories. The quotient theory is defined by orbifolded theory $\mathbf{D}_{i}=\mathbf{B}_{i}/Z_{N_{i}/\text{gcd}(N_{1},N_{2})}$ (or parity zero sector of $\mathbf{F}_{i}/Z_{N_{i}/\text{gcd}(N_{1},N_{2})}$) corresponding to the red arrows in the figure. By applying the folding trick (black arrow) and quotient operation to $\mathbf{B}_{1} \otimes \overline{\mathbf{B}_{2}}$, one will obtain the (homo)morphism between $\mathbf{D}_{1}$ and $\mathbf{D}_{2}$. The new partition function and the corresponding extended theory $\mathbf{F}_{1\otimes 2}$ (blue arrow) will provide the algebraic data of the conformal interfaces. One can see the dual relationship between greatest common divisor and lowest common multiple as a dual relationship between quotient and extension. Hence, for further understanding of the quotient or gauging structure, it is necessary to study the extension of theories more thoroughly.}
\label{gcd_lcm}
\end{center}
\end{figure}

We stress that the above quotient by subgroup structure is exact and rigorous, and this can be considered as a natural generalization of group quotient in a fusion ring. In other words, after sufficient extensions, the definition of a quotient will become exact even when the straightforward definition is ill-defined. Corresponding discussions appear ubiquitously in the literature of orbifolding.

The same mismatch between the preserved symmetry and the generator of extension of coupled models appears in studying recent noninvertible (or nonabelian) symmetry preserving RG flows\cite{Nakayama:2024msv}. As a simplest example, in the massless RG flow from the tricritical Ising model to the Ising model, the Ising fusion ring symmetry generated by three objects $\{ I,\psi, \sigma \}$ survives under the massless flow, where $\{ I, \psi\}$ is the $Z_{2}$ group and $\sigma$ is the Ising anyon satisfying the relations $\sigma \times \sigma =I+\psi$ and $\sigma \times \psi=\sigma$. However, the extension generating the coupled model is the common $Z_{2}$ group, $\{ I, \psi\}$. In other words, the $Z_{2}$ group is a trivial common quotient group. The precise relationship between the discussions in this subsection and gauging by noninvertible symmetries in recent literature\cite{Lu:2022ver,Perez-Lona:2023djo,Choi:2023vgk,Diatlyk:2023fwf,Perez-Lona:2024sds,Lu:2024ytl,Ando:2024hun,Lu:2024lzf,Yu:2025iqf,Maeda:2025rxc,Lu:2025gpt,Seifnashri:2025fgd,Tanaka:2025qou} is an interesting future problem.

\section{Conclusion}
\label{conclusion}

In this work, we have provided a general method to extend a CFT by its group symmetry. By using the terminology in category theory, our construction provides the method to obtain algebraic data of a $Z_{N}$ symmetry graded fusion category and $Z_{N}$ graded SymTFT systematically. There exist similar proposals to obtain a $Z_{N}$ (or a general group $G$) graded category\cite{etingof2009weaklygrouptheoreticalsolvablefusion,Galindo:2024qzg}. However, most of the existing literature lacks the analysis of $Z_{N}$ symmetry action on the original category theory, and its anomaly-free conditions are difficult to read for researchers in other fields. We stress the significance of integer spin simple currents again for the construction of TQFTs and TOs.

We remark on an open problem for the interpretation of the new series of extended models constructed by coupling multiple copies of CFTs. As the author has clarified in \cite{Fukusumi:2024ejk}, the extension of models is closely connected to symmetry-preserving domain walls. In this setting, the common subgroup of $\text{gcd}(\{N_{i_{+}}, N_{i_{-}} \})$ symmetry seems to be preserved. Interestingly, the preserved symmetry and the symmetry that generates extensions are different.

When studying $SU(N)_{K}$ WZW models, the classification by $Z_{N}$ symmetry under fixed $N$ has already been studied widely \cite{Haldane:1981zza,Haldane:1982rj,Haldane:1983ru,Affleck:1988wz,Lecheminant:2015iga,Furuya:2015coa,Numasawa:2017crf,Yao:2018kel,Wamer:2019oge,Fukusumi_2022_c,Kikuchi:2022ipr,Herviou:2023unm,Lecheminant:2024apm,Fukusumi:2024ejk}. As a natural generalization, we propose that the quotient group and its anomaly structure govern such classifications by varying both $N$ and $K$. By applying CFT/TQFT correspondence to this phenomenology, one can also obtain the corresponding classification of $2+1$ dimensional spin liquid based on its quotient group. The significance of orbifolding or gauging can be seen commonly in condensed matter literature\cite{Levin_2012,Sule:2013qla,Hsieh:2014lba,Chen_2017_2,Tiwari:2017wqf,Li:2023mmw,Li:2023knf,Chatterjee:2024ych}. For readers interested in the application of orbifolding in condensed matter, we note the review \cite{Chiu:2015mfr}. Whereas we have restricted our attention to the system graded by the $Z_{N}$ group by the extensions, one can study corresponding systems graded by a ring. For example, the tricritical Ising fusion ring is isomorphic to the tensor decomposition (Fibonacci)$\otimes$(Ising), and this fusion ring can be graded by the Ising fusion ring. Hence, with some proper extensions, domain wall preserving quotient ring grading will appear in general. Application of this argument to higher-dimensional systems is also an interesting problem.

\section{Acknowledgement}
YF thanks Jurgen Fuchs and Yuji Tachikawa for the helpful discussions on the relation between fusion rings and category theory, Taishi Kawamoto for the discussion on noninvertible symmetry, Zhengdi Sun for the helpful discussions on gauging operations, Yuma Furuta for the fruitful collaboration in the related project, and Yifan Liu for the discussions on models with nonanomalous subgroups. We thank Kohki Kawabata for the discussion on discrete symmetry on related models, and Takuya Okuda, Weiguang Cao, Xingyang Yu, Yamato Honda, and Zhian Jia for the helpful comments on the manuscript. YF thanks the support of NCTS. The work of SY was supported by FoPM, WINGS Program, the University of Tokyo.

\appendix

\section{Twisted fusion rings: Generalization of even-odd problems}
\label{twisted_fusion}

In this section, we show a procedure for obtaining a fusion ring twisted by chiral (or antichiral) $Z_{N}$ operations. Because of the twist, the coefficients of the fusion ring can be complex numbers. This phenomenon might be unfamiliar to the readers, but a nonpositive noninteger fusion coefficient has already appeared in recent literature. Moreover, the difference between twisted and untwisted theories appears as an even-odd problem in one-dimensional quantum systems\cite{Li:2023knf,Li:2023mmw}. We also note that if we interpret the fusion ring as a boundary CFT, this kind of nonpositive integer coefficient should be prohibited \cite{Cardy:1989ir}. However, if we interpret it as an operator algebra formed by a topological operator (or SymTFT), such an appearance of an unusual fusion ring will be possible in principle. We remind the readers of one of the most fundamental facts that the set of operators or matrices in a system, such as in lattice models and quantum field theories, is a ring over a complex number field. The set of integers does not form a ring, and it is less general and more difficult to capture the property of the symmetry operator. 

We use the notation in the main text and introduce the $Z_{N}$ chiral parity operator, which acts on the chiral objects as follows,
\begin{equation}
\omega^{F} \phi_{a,p,q}^{(n)}=\omega^{nq+p} \phi_{a,p,q}^{(n)}
\end{equation}
where the number $\omega=e^{\frac{2 \pi i}{N}}$ is the $N$-th root of unity. By introducing this chiral parity operator, one can introduce the following twisted chiral theory $\mathbf{F}^{[l]}$,
\begin{equation}
\Phi_{a,p,\overline{p},q}^{(n),[l]}=\omega^{lF} \Phi_{a,p,\overline{p},q}^{(n)}=\sum_{q'=0}^{\frac{N}{n}-1} \frac{\omega^{l(nq'+p)} \phi^{(n)}_{a,p,q'}\otimes_{\text{D}}\overline{\phi}^{(n)}_{a,\overline{p},q-q'}}{\sqrt{N/n}}
\end{equation}
Corresponding to the number $l$, the twisted fusion rule will appear. By replacing the symbol $\Phi$ to $\Phi^{[l]}$, one can obtain the corresponding SymTFT $\mathbf{S}^{[l]}=\{ \Psi^{[l]}\}$ or $\{\Theta^{[l]}\}$. The periodicity coming from $l$ modulo $N$ will be realized in site number dependence in quantum lattice models, and this can be considered as a generalization of the even-odd problem\cite{Hagendorf_2012,Hagendorf:2012fz} (See also the introduction of \cite{Matsui:2016oqq}). Moreover, this type of periodicity plays a role in performing the duality transformation in a quantum spin chain\cite{Li:2023knf,Li:2023mmw}. We also note works studying unusual periodicities in $Z_{N}$ symmetric quantum many-body systems\cite{Watanabe:2021wwt,Hu_2023}.





\bibliography{subgroup_extension.bib}

\begin{thebibliography}{100}
\providecommand{\url}[1]{\texttt{#1}}
\providecommand{\urlprefix}{URL }
\expandafter\ifx\csname urlstyle\endcsname\relax
  \providecommand{\doi}[1]{doi:\discretionary{}{}{}#1}\else
  \providecommand{\doi}{doi:\discretionary{}{}{}\begingroup
  \urlstyle{rm}\Url}\fi
\providecommand{\eprint}[2][]{\url{#2}}

\bibitem{Slansky:1981yr}
R.~Slansky,
\newblock \emph{{Group Theory for Unified Model Building}},
\newblock Phys. Rept. \textbf{79}, 1 (1981),
\newblock \doi{10.1016/0370-1573(81)90092-2}.

\bibitem{tHooft:1979rat}
G.~'t~Hooft,
\newblock \emph{{Naturalness, chiral symmetry, and spontaneous chiral symmetry
  breaking}},
\newblock NATO Sci. Ser. B \textbf{59}, 135 (1980),
\newblock \doi{10.1007/978-1-4684-7571-5_9}.

\bibitem{Lieb:1961fr}
E.~H. Lieb, T.~Schultz and D.~Mattis,
\newblock \emph{{Two soluble models of an antiferromagnetic chain}},
\newblock Annals Phys. \textbf{16}, 407 (1961),
\newblock \doi{10.1016/0003-4916(61)90115-4}.

\bibitem{Gaiotto:2014kfa}
D.~Gaiotto, A.~Kapustin, N.~Seiberg and B.~Willett,
\newblock \emph{{Generalized Global Symmetries}},
\newblock JHEP \textbf{02}, 172 (2015),
\newblock \doi{10.1007/JHEP02(2015)172},
\newblock \eprint{1412.5148}.

\bibitem{Petkova:2000ip}
V.~B. Petkova and J.~B. Zuber,
\newblock \emph{{Generalized Twisted Partition Functions}},
\newblock Phys. Lett. B \textbf{504}, 157 (2001),
\newblock \doi{10.1016/S0370-2693(01)00276-3},
\newblock \eprint{hep-th/0011021}.

\bibitem{Graham:2003nc}
K.~Graham and G.~M.~T. Watts,
\newblock \emph{{Defect lines and boundary flows}},
\newblock JHEP \textbf{04}, 019 (2004),
\newblock \doi{10.1088/1126-6708/2004/04/019},
\newblock \eprint{hep-th/0306167}.

\bibitem{Frohlich:2004ef}
J.~Frohlich, J.~Fuchs, I.~Runkel and C.~Schweigert,
\newblock \emph{{Kramers-Wannier duality from conformal defects}},
\newblock Phys. Rev. Lett. \textbf{93}, 070601 (2004),
\newblock \doi{10.1103/PhysRevLett.93.070601},
\newblock \eprint{cond-mat/0404051}.

\bibitem{Frohlich:2006ch}
J.~Frohlich, J.~Fuchs, I.~Runkel and C.~Schweigert,
\newblock \emph{{Duality and defects in rational conformal field theory}},
\newblock Nucl. Phys. B \textbf{763}, 354 (2007),
\newblock \doi{10.1016/j.nuclphysb.2006.11.017},
\newblock \eprint{hep-th/0607247}.

\bibitem{Cobanera:2009as}
E.~Cobanera, G.~Ortiz and Z.~Nussinov,
\newblock \emph{{Unified approach to Quantum and Classical Dualities}},
\newblock Phys. Rev. Lett. \textbf{104}, 020402 (2010),
\newblock \doi{10.1103/PhysRevLett.104.020402},
\newblock \eprint{0907.0733}.

\bibitem{Cobanera:2011wn}
E.~Cobanera, G.~Ortiz and Z.~Nussinov,
\newblock \emph{{The Bond-Algebraic Approach to Dualities}},
\newblock Adv. Phys. \textbf{60}, 679 (2011),
\newblock \doi{10.1080/00018732.2011.619814},
\newblock \eprint{1103.2776}.

\bibitem{Cobanera:2012dc}
E.~Cobanera, G.~Ortiz and Z.~Nussinov,
\newblock \emph{{Holographic symmetries and generalized order parameters for
  topological matter}},
\newblock Phys. Rev. B \textbf{87}(4), 041105 (2013),
\newblock \doi{10.1103/PhysRevB.87.041105},
\newblock \eprint{1211.0564}.

\bibitem{article}
A.~Ocneanu,
\newblock \emph{Paths on coxeter diagrams: From platonic solids and
  singularities to minimal models and subfactors},
\newblock Lectures on Operator Theory  (2000).

\bibitem{Bockenhauer:1999wt}
J.~Bockenhauer, D.~E. Evans and Y.~Kawahigashi,
\newblock \emph{{Chiral structure of modular invariants for subfactors}},
\newblock Commun. Math. Phys. \textbf{210}, 733 (2000),
\newblock \doi{10.1007/s002200050798},
\newblock \eprint{math/9907149}.

\bibitem{Kawahigashi:2021hds}
Y.~Kawahigashi,
\newblock \emph{{Two-dimensional topological order and operator algebras}},
\newblock Int. J. Mod. Phys. B \textbf{35}(08), 2130003 (2021),
\newblock \doi{10.1142/S0217979221300036},
\newblock \eprint{2102.10953}.

\bibitem{Evans:2023nbp}
D.~E. Evans and Y.~Kawahigashi,
\newblock \emph{{Subfactors and mathematical physics}},
\newblock Bull. Am. Math. Soc. \textbf{60}(4), 459 (2023),
\newblock \doi{10.1090/bull/1799},
\newblock \eprint{2303.04459}.

\bibitem{Harvey:2005it}
J.~A. Harvey,
\newblock \emph{{TASI 2003 lectures on anomalies}} (2005),
  \eprint{hep-th/0509097}.

\bibitem{McGreevy:2022oyu}
J.~McGreevy,
\newblock \emph{{Generalized Symmetries in Condensed Matter}},
\newblock Ann. Rev. Condensed Matter Phys. \textbf{14}, 57 (2023),
\newblock \doi{10.1146/annurev-conmatphys-040721-021029},
\newblock \eprint{2204.03045}.

\bibitem{Cordova:2022ruw}
C.~Cordova, T.~T. Dumitrescu, K.~Intriligator and S.-H. Shao,
\newblock \emph{{Snowmass White Paper: Generalized Symmetries in Quantum Field
  Theory and Beyond}},
\newblock In \emph{{Snowmass 2021}} (2022), \eprint{2205.09545}.

\bibitem{Bhardwaj:2023kri}
L.~Bhardwaj, L.~E. Bottini, L.~Fraser-Taliente, L.~Gladden, D.~S.~W. Gould,
  A.~Platschorre and H.~Tillim,
\newblock \emph{{Lectures on generalized symmetries}},
\newblock Phys. Rept. \textbf{1051}, 1 (2024),
\newblock \doi{10.1016/j.physrep.2023.11.002},
\newblock \eprint{2307.07547}.

\bibitem{Brennan:2023mmt}
T.~D. Brennan and S.~Hong,
\newblock \emph{{Introduction to Generalized Global Symmetries in QFT and
  Particle Physics}}  (2023),
\newblock \eprint{2306.00912}.

\bibitem{Cappelli:1987xt}
A.~Cappelli, C.~Itzykson and J.~B. Zuber,
\newblock \emph{{The ADE Classification of Minimal andA \textbf{1} (1)
  Conformal Invariant Theories}},
\newblock Commun. Math. Phys. \textbf{113}, 1 (1987),
\newblock \doi{10.1007/BF01221394}.

\bibitem{Vafa:1986wx}
C.~Vafa,
\newblock \emph{{Modular Invariance and Discrete Torsion on Orbifolds}},
\newblock Nucl. Phys. B \textbf{273}, 592 (1986),
\newblock \doi{10.1016/0550-3213(86)90379-2}.

\bibitem{Dixon:1986jc}
L.~J. Dixon, J.~A. Harvey, C.~Vafa and E.~Witten,
\newblock \emph{{Strings on Orbifolds. 2.}},
\newblock Nucl. Phys. B \textbf{274}, 285 (1986),
\newblock \doi{10.1016/0550-3213(86)90287-7}.

\bibitem{Schellekens:1989uf}
A.~N. Schellekens and S.~Yankielowicz,
\newblock \emph{{Field Identification Fixed Points in the Coset Construction}},
\newblock Nucl. Phys. B \textbf{334}, 67 (1990),
\newblock \doi{10.1016/0550-3213(90)90657-Y}.

\bibitem{Gato-Rivera:1990lxi}
B.~Gato-Rivera and A.~N. Schellekens,
\newblock \emph{{Complete classification of simple current automorphisms}},
\newblock Nucl. Phys. B \textbf{353}, 519 (1991),
\newblock \doi{10.1016/0550-3213(91)90346-Y}.

\bibitem{Schellekens:1990ys}
A.~N. Schellekens,
\newblock \emph{{Fusion rule automorphisms from integer spin simple currents}},
\newblock Phys. Lett. B \textbf{244}, 255 (1990),
\newblock \doi{10.1016/0370-2693(90)90065-E}.

\bibitem{Gato-Rivera:1991bcq}
B.~Gato-Rivera and A.~N. Schellekens,
\newblock \emph{{Classification of simple current invariants}},
\newblock In \emph{{Joint International Lepton Photon Symposium at High
  Energies (15th) and European Physical Society Conference on High-energy
  Physics}} (1991), \eprint{hep-th/9109035}.

\bibitem{Gato-Rivera:1991bqv}
B.~Gato-Rivera and A.~N. Schellekens,
\newblock \emph{{Complete classification of simple current modular invariants
  for (Z(p))**k}},
\newblock Commun. Math. Phys. \textbf{145}, 85 (1992),
\newblock \doi{10.1007/BF02099282}.

\bibitem{Kreuzer:1993tf}
M.~Kreuzer and A.~N. Schellekens,
\newblock \emph{{Simple currents versus orbifolds with discrete torsion: A
  Complete classification}},
\newblock Nucl. Phys. B \textbf{411}, 97 (1994),
\newblock \doi{10.1016/0550-3213(94)90055-8},
\newblock \eprint{hep-th/9306145}.

\bibitem{Fuchs:1996rq}
J.~Fuchs, B.~Schellekens and C.~Schweigert,
\newblock \emph{{Fixed Point Resolution in Conformal Field Theory}},
\newblock In \emph{{21st International Colloquium on Group Theoretical Methods
  in Physics}} (1996), \eprint{hep-th/9612093}.

\bibitem{Fuchs:1996dd}
J.~Fuchs, A.~N. Schellekens and C.~Schweigert,
\newblock \emph{{A Matrix S for All Simple Current Extensions}},
\newblock Nucl. Phys. B \textbf{473}, 323 (1996),
\newblock \doi{10.1016/0550-3213(96)00247-7},
\newblock \eprint{hep-th/9601078}.

\bibitem{Witten:1988hf}
E.~Witten,
\newblock \emph{{Quantum Field Theory and the Jones Polynomial}},
\newblock Commun. Math. Phys. \textbf{121}, 351 (1989),
\newblock \doi{10.1007/BF01217730}.

\bibitem{Laughlin:1983fy}
R.~B. Laughlin,
\newblock \emph{{Anomalous quantum Hall effect: An Incompressible quantum fluid
  with fractionallycharged excitations}},
\newblock Phys. Rev. Lett. \textbf{50}, 1395 (1983),
\newblock \doi{10.1103/PhysRevLett.50.1395}.

\bibitem{Blok:1991zq}
B.~Blok and X.~G. Wen,
\newblock \emph{{Many body systems with nonAbelian statistics}},
\newblock Nucl. Phys. B \textbf{374}, 615 (1992),
\newblock \doi{10.1016/0550-3213(92)90402-W}.

\bibitem{Read:1998ed}
N.~Read and E.~Rezayi,
\newblock \emph{{Beyond paired quantum Hall states: Parafermions and
  incompressible states in the first excited Landau level}},
\newblock Phys. Rev. B \textbf{59}, 8084 (1999),
\newblock \doi{10.1103/PhysRevB.59.8084},
\newblock \eprint{cond-mat/9809384}.

\bibitem{Schoutens:2015uia}
K.~Schoutens and X.-G. Wen,
\newblock \emph{{Simple-current algebra constructions of 2+1-dimensional
  topological orders}},
\newblock Phys. Rev. B \textbf{93}(4), 045109 (2016),
\newblock \doi{10.1103/PhysRevB.93.045109},
\newblock \eprint{1508.01111}.

\bibitem{Dorey:2016mxm}
N.~Dorey, D.~Tong and C.~Turner,
\newblock \emph{{Matrix model for non-Abelian quantum Hall states}},
\newblock Phys. Rev. B \textbf{94}(8), 085114 (2016),
\newblock \doi{10.1103/PhysRevB.94.085114},
\newblock \eprint{1603.09688}.

\bibitem{Fuji_2017}
Y.~Fuji and P.~Lecheminant,
\newblock Physical Review B \textbf{95}(12) (2017),
\newblock \doi{10.1103/physrevb.95.125130}.

\bibitem{Bourgine:2024ycr}
J.-E. Bourgine and Y.~Matsuo,
\newblock \emph{{Calogero model for the non-Abelian quantum Hall effect}},
\newblock Phys. Rev. B \textbf{109}(15), 155158 (2024),
\newblock \doi{10.1103/PhysRevB.109.155158},
\newblock \eprint{2401.03087}.

\bibitem{Fukusumi:2023psx}
Y.~Fukusumi, G.~Ji and B.~Yang,
\newblock \emph{{Operator-state correspondence in simple current extended
  conformal field theories: Toward a general understanding of chiral conformal
  field theories and topological orders}}  (2023),
\newblock \eprint{2311.15621}.

\bibitem{Gliozzi:1976qd}
F.~Gliozzi, J.~Scherk and D.~I. Olive,
\newblock \emph{{Supersymmetry, Supergravity Theories and the Dual Spinor
  Model}},
\newblock Nucl. Phys. B \textbf{122}, 253 (1977),
\newblock \doi{10.1016/0550-3213(77)90206-1}.

\bibitem{Fukusumi:2022xxe}
Y.~Fukusumi and B.~Yang,
\newblock \emph{{Fermionic fractional quantum Hall states: A modern approach to
  systems with bulk-edge correspondence}},
\newblock Phys. Rev. B \textbf{108}(8), 085123 (2023),
\newblock \doi{10.1103/PhysRevB.108.085123},
\newblock \eprint{2212.12993}.

\bibitem{Fukusumi_2022_c}
Y.~Fukusumi,
\newblock \emph{{Composing parafermions: a construction of $Z_{N}$ fractional
  quantum Hall systems and a modern understanding of confinement and duality}}
  (2022),
\newblock \eprint{2212.12999}.

\bibitem{turaev2000homotopyfieldtheorydimension}
V.~Turaev,
\newblock \emph{Homotopy field theory in dimension 3 and crossed
  group-categories} (2000), \eprint{math/0005291}.

\bibitem{kirillov2001modularcategoriesorbifoldmodels}
A.~K. Jr,
\newblock \emph{Modular categories and orbifold models ii} (2001),
  \eprint{math/0110221}.

\bibitem{Frohlich:2003hm}
J.~Frohlich, J.~Fuchs, I.~Runkel and C.~Schweigert,
\newblock \emph{{Correspondences of ribbon categories}},
\newblock Adv. Math. \textbf{199}, 192 (2006),
\newblock \doi{10.1016/j.aim.2005.04.007},
\newblock \eprint{math/0309465}.

\bibitem{mueger2004galoisextensionsbraidedtensor}
M.~Mueger,
\newblock \emph{Galois extensions of braided tensor categories and braided
  crossed g-categories} (2004), \eprint{math/0209093}.

\bibitem{etingof2009weaklygrouptheoreticalsolvablefusion}
P.~Etingof, D.~Nikshych and V.~Ostrik,
\newblock \emph{Weakly group-theoretical and solvable fusion categories}
  (2009), \eprint{0809.3031}.

\bibitem{Hamidi:1986vh}
S.~Hamidi and C.~Vafa,
\newblock \emph{{Interactions on Orbifolds}},
\newblock Nucl. Phys. B \textbf{279}, 465 (1987),
\newblock \doi{10.1016/0550-3213(87)90006-X}.

\bibitem{Dijkgraaf:1989hb}
R.~Dijkgraaf, C.~Vafa, E.~P. Verlinde and H.~L. Verlinde,
\newblock \emph{{The Operator Algebra of Orbifold Models}},
\newblock Commun. Math. Phys. \textbf{123}, 485 (1989),
\newblock \doi{10.1007/BF01238812}.

\bibitem{Vafa:1989ih}
C.~Vafa,
\newblock \emph{{Quantum Symmetries of String Vacua}},
\newblock Mod. Phys. Lett. A \textbf{4}, 1615 (1989),
\newblock \doi{10.1142/S0217732389001842}.

\bibitem{Tachikawa:2017gyf}
Y.~Tachikawa,
\newblock \emph{{On gauging finite subgroups}},
\newblock SciPost Phys. \textbf{8}(1), 015 (2020),
\newblock \doi{10.21468/SciPostPhys.8.1.015},
\newblock \eprint{1712.09542}.

\bibitem{Bhardwaj:2017xup}
L.~Bhardwaj and Y.~Tachikawa,
\newblock \emph{{On finite symmetries and their gauging in two dimensions}},
\newblock JHEP \textbf{03}, 189 (2018),
\newblock \doi{10.1007/JHEP03(2018)189},
\newblock \eprint{1704.02330}.

\bibitem{Ginsparg:1988ui}
P.~H. Ginsparg,
\newblock \emph{{APPLIED CONFORMAL FIELD THEORY}},
\newblock In \emph{{Les Houches Summer School in Theoretical Physics: Fields,
  Strings, Critical Phenomena}} (1988), \eprint{hep-th/9108028}.

\bibitem{Moore:1991ks}
G.~W. Moore and N.~Read,
\newblock \emph{{Nonabelions in the fractional quantum Hall effect}},
\newblock Nucl. Phys. B \textbf{360}, 362 (1991),
\newblock \doi{10.1016/0550-3213(91)90407-O}.

\bibitem{Cappelli:1996np}
A.~Cappelli and G.~R. Zemba,
\newblock \emph{{Modular invariant partition functions in the quantum Hall
  effect}},
\newblock Nucl. Phys. B \textbf{490}, 595 (1997),
\newblock \doi{10.1016/S0550-3213(97)00110-7},
\newblock \eprint{hep-th/9605127}.

\bibitem{Milovanovic:1996nj}
M.~Milovanovic and N.~Read,
\newblock \emph{{Edge excitations of paired fractional quantum Hall states}},
\newblock Phys. Rev. B \textbf{53}, 13559 (1996),
\newblock \doi{10.1103/PhysRevB.53.13559},
\newblock \eprint{cond-mat/9602113}.

\bibitem{Cappelli:2010jv}
A.~Cappelli and G.~Viola,
\newblock \emph{{Partition Functions of Non-Abelian Quantum Hall States}},
\newblock J. Phys. A \textbf{44}, 075401 (2011),
\newblock \doi{10.1088/1751-8113/44/7/075401},
\newblock \eprint{1007.1732}.

\bibitem{Ino:1998by}
K.~Ino,
\newblock \emph{{Modular invariants in the fractional quantum Hall effect}},
\newblock Nucl. Phys. B \textbf{532}, 783 (1998),
\newblock \doi{10.1016/S0550-3213(98)00598-7},
\newblock \eprint{cond-mat/9804198}.

\bibitem{Petkova:1988cy}
V.~B. Petkova,
\newblock \emph{{Two-dimensional (Half) Integer Spin Conformal Theories With
  Central Charge $C < 1$}},
\newblock Int. J. Mod. Phys. A \textbf{3}, 2945 (1988),
\newblock \doi{10.1142/S0217751X88001235}.

\bibitem{Furlan:1989ra}
P.~Furlan, A.~C. Ganchev and V.~B. Petkova,
\newblock \emph{{Fusion Matrices and $C < 1$ (Quasi)local Conformal Theories}},
\newblock Int. J. Mod. Phys. A \textbf{5}, 2721 (1990),
\newblock \doi{10.1142/S0217751X90001252},
\newblock [Erratum: Int.J.Mod.Phys.A 5, 3641 (1990)].

\bibitem{Runkel:2020zgg}
I.~Runkel and G.~M.~T. Watts,
\newblock \emph{{Fermionic CFTs and classifying algebras}},
\newblock JHEP \textbf{06}, 025 (2020),
\newblock \doi{10.1007/JHEP06(2020)025},
\newblock \eprint{2001.05055}.

\bibitem{Hsieh:2020uwb}
C.-T. Hsieh, Y.~Nakayama and Y.~Tachikawa,
\newblock \emph{{Fermionic minimal models}},
\newblock Phys. Rev. Lett. \textbf{126}(19), 195701 (2021),
\newblock \doi{10.1103/PhysRevLett.126.195701},
\newblock \eprint{2002.12283}.

\bibitem{Nielsen:1980rz}
H.~B. Nielsen and M.~Ninomiya,
\newblock \emph{{Absence of Neutrinos on a Lattice. 1. Proof by Homotopy
  Theory}},
\newblock Nucl. Phys. B \textbf{185}, 20 (1981),
\newblock \doi{10.1016/0550-3213(82)90011-6},
\newblock [Erratum: Nucl.Phys.B 195, 541 (1982)].

\bibitem{Nielsen:1981xu}
H.~B. Nielsen and M.~Ninomiya,
\newblock \emph{{Absence of Neutrinos on a Lattice. 2. Intuitive Topological
  Proof}},
\newblock Nucl. Phys. B \textbf{193}, 173 (1981),
\newblock \doi{10.1016/0550-3213(81)90524-1}.

\bibitem{Goddard:1984vk}
P.~Goddard, A.~Kent and D.~I. Olive,
\newblock \emph{{Virasoro Algebras and Coset Space Models}},
\newblock Phys. Lett. B \textbf{152}, 88 (1985),
\newblock \doi{10.1016/0370-2693(85)91145-1}.

\bibitem{Goddard:1984hg}
P.~Goddard and D.~I. Olive,
\newblock \emph{{Kac-Moody Algebras, Conformal Symmetry and Critical
  Exponents}},
\newblock Nucl. Phys. B \textbf{257}, 226 (1985),
\newblock \doi{10.1016/0550-3213(85)90344-X}.

\bibitem{Fukusumi:2024ejk}
Y.~Fukusumi,
\newblock \emph{{Gauging or extending bulk and boundary conformal field
  theories: Application to bulk and domain wall problem in topological matter
  and their descriptions by (mock) modular covariant}}  (2024),
\newblock \eprint{2412.19577}.

\bibitem{Kitaev:2006lla}
A.~Kitaev,
\newblock \emph{{Anyons in an exactly solved model and beyond}},
\newblock Annals Phys. \textbf{321}(1), 2 (2006),
\newblock \doi{10.1016/j.aop.2005.10.005},
\newblock \eprint{cond-mat/0506438}.

\bibitem{Gaiotto:2015zta}
D.~Gaiotto and A.~Kapustin,
\newblock \emph{{Spin TQFTs and fermionic phases of matter}},
\newblock Int. J. Mod. Phys. A \textbf{31}(28n29), 1645044 (2016),
\newblock \doi{10.1142/S0217751X16450445},
\newblock \eprint{1505.05856}.

\bibitem{Lan_2016}
T.~Lan, L.~Kong and X.-G. Wen,
\newblock \emph{Theory of ($2+1$)-dimensional fermionic topological orders and
  fermionic/bosonic topological orders with symmetries},
\newblock Physical Review B \textbf{94}(15) (2016),
\newblock \doi{10.1103/physrevb.94.155113}.

\bibitem{Lan2016ModularEO}
T.~Lan, L.~Kong and X.-G. Wen,
\newblock \emph{Modular extensions of unitary braided fusion categories and
  2+1d topological/spt orders with symmetries},
\newblock Communications in Mathematical Physics \textbf{351}, 709  (2016).

\bibitem{Cho:2022kzf}
G.~Y. Cho, H.-c. Kim, D.~Seo and M.~You,
\newblock \emph{{Classification of fermionic topological orders from congruence
  representations}},
\newblock Phys. Rev. B \textbf{108}(11), 115103 (2023),
\newblock \doi{10.1103/PhysRevB.108.115103},
\newblock \eprint{2210.03681}.

\bibitem{Goddard:1986ee}
P.~Goddard, A.~Kent and D.~I. Olive,
\newblock \emph{{Unitary Representations of the Virasoro and Supervirasoro
  Algebras}},
\newblock Commun. Math. Phys. \textbf{103}, 105 (1986),
\newblock \doi{10.1007/BF01464283}.

\bibitem{Andrews:1984af}
G.~E. Andrews, R.~J. Baxter and P.~J. Forrester,
\newblock \emph{{Eight vertex SOS model and generalized Rogers-Ramanujan type
  identities}},
\newblock J. Statist. Phys. \textbf{35}, 193 (1984),
\newblock \doi{10.1007/BF01014383}.

\bibitem{Kedem:1993ze}
R.~Kedem, T.~R. Klassen, B.~M. McCoy and E.~Melzer,
\newblock \emph{{Fermionic sum representations for conformal field theory
  characters}},
\newblock Phys. Lett. B \textbf{307}, 68 (1993),
\newblock \doi{10.1016/0370-2693(93)90194-M},
\newblock \eprint{hep-th/9301046}.

\bibitem{Campbell_2024}
G.~B. Campbell,
\newblock \emph{Vector Partitions, Visible Points and Ramanujan Functions},
\newblock Chapman and Hall/CRC,
\newblock ISBN 9781003174158,
\newblock \doi{10.1201/9781003174158} (2024).

\bibitem{Harvey:2019htf}
J.~A. Harvey,
\newblock \emph{{Ramanujan's influence on string theory, black holes and
  moonshine}}  (2019),
\newblock \doi{10.1098/rsta.2018.0440},
\newblock \eprint{1909.11477}.

\bibitem{Fukusumi:2024cnl}
Y.~Fukusumi,
\newblock \emph{{Fusion rule in conformal field theories and topological
  orders: A unified view of correspondence and (fractional) supersymmetry and
  their relation to topological holography}}  (2024),
\newblock \eprint{2405.05178}.

\bibitem{Brunner:2007ur}
I.~Brunner and D.~Roggenkamp,
\newblock \emph{{Defects and bulk perturbations of boundary Landau-Ginzburg
  orbifolds}},
\newblock JHEP \textbf{04}, 001 (2008),
\newblock \doi{10.1088/1126-6708/2008/04/001},
\newblock \eprint{0712.0188}.

\bibitem{Gaiotto:2012np}
D.~Gaiotto,
\newblock \emph{{Domain Walls for Two-Dimensional Renormalization Group
  Flows}},
\newblock JHEP \textbf{12}, 103 (2012),
\newblock \doi{10.1007/JHEP12(2012)103},
\newblock \eprint{1201.0767}.

\bibitem{Fukusumi:2025clr}
Y.~Fukusumi and Y.~Furuta,
\newblock \emph{{Homomorphism, substructure and ideal: Elementary but rigorous
  aspects of renormalization group or hierarchical structure of topological
  orders}}  (2025),
\newblock \eprint{2506.23155}.

\bibitem{Sule:2013qla}
O.~M. Sule, X.~Chen and S.~Ryu,
\newblock \emph{{Symmetry-protected topological phases and orbifolds:
  Generalized Laughlin's argument}},
\newblock Phys. Rev. B \textbf{88}, 075125 (2013),
\newblock \doi{10.1103/PhysRevB.88.075125},
\newblock \eprint{1305.0700}.

\bibitem{Hsieh:2014lba}
C.-T. Hsieh, O.~M. Sule, G.~Y. Cho, S.~Ryu and R.~G. Leigh,
\newblock \emph{{Symmetry-protected Topological Phases, Generalized Laughlin
  Argument and Orientifolds}},
\newblock Phys. Rev. B \textbf{90}(16), 165134 (2014),
\newblock \doi{10.1103/PhysRevB.90.165134},
\newblock \eprint{1403.6902}.

\bibitem{Chen_2017_2}
X.~Chen, A.~Roy, J.~C.~Y. Teo and S.~Ryu,
\newblock \emph{From orbifolding conformal field theories to gauging
  topological phases},
\newblock Physical Review B \textbf{96}(11) (2017),
\newblock \doi{10.1103/physrevb.96.115447}.

\bibitem{Tiwari:2017wqf}
A.~Tiwari, X.~Chen, K.~Shiozaki and S.~Ryu,
\newblock \emph{{Bosonic topological phases of matter: Bulk-boundary
  correspondence, symmetry protected topological invariants, and gauging}},
\newblock Phys. Rev. B \textbf{97}(24), 245133 (2018),
\newblock \doi{10.1103/PhysRevB.97.245133},
\newblock \eprint{1710.04730}.

\bibitem{Affleck:1998nq}
I.~Affleck, M.~Oshikawa and H.~Saleur,
\newblock \emph{{Boundary critical phenomena in the three state Potts model}},
\newblock J. Phys. A \textbf{31}, 5827 (1998),
\newblock \doi{10.1088/0305-4470/31/28/003},
\newblock \eprint{cond-mat/9804117}.

\bibitem{Fuchs:1999zi}
J.~Fuchs and C.~Schweigert,
\newblock \emph{{Symmetry breaking boundaries. 1. General theory}},
\newblock Nucl. Phys. B \textbf{558}, 419 (1999),
\newblock \doi{10.1016/S0550-3213(99)00406-X},
\newblock \eprint{hep-th/9902132}.

\bibitem{Birke:1999ik}
L.~Birke, J.~Fuchs and C.~Schweigert,
\newblock \emph{{Symmetry breaking boundary conditions and WZW orbifolds}},
\newblock Adv. Theor. Math. Phys. \textbf{3}, 671 (1999),
\newblock \doi{10.4310/ATMP.1999.v3.n3.a8},
\newblock \eprint{hep-th/9905038}.

\bibitem{Quella:2002ct}
T.~Quella and V.~Schomerus,
\newblock \emph{{Symmetry breaking boundary states and defect lines}},
\newblock JHEP \textbf{06}, 028 (2002),
\newblock \doi{10.1088/1126-6708/2002/06/028},
\newblock \eprint{hep-th/0203161}.

\bibitem{Fukusumi_2022}
Y.~Fukusumi,
\newblock \emph{{Gaplessness protected by bulk-edge correspondence}}  (2022),
\newblock \eprint{2212.12996}.

\bibitem{Lin:2022dhv}
Y.-H. Lin, M.~Okada, S.~Seifnashri and Y.~Tachikawa,
\newblock \emph{{Asymptotic density of states in 2d CFTs with non-invertible
  symmetries}},
\newblock JHEP \textbf{03}, 094 (2023),
\newblock \doi{10.1007/JHEP03(2023)094},
\newblock \eprint{2208.05495}.

\bibitem{Yao:2020dqx}
Y.~Yao and A.~Furusaki,
\newblock \emph{{Parafermionization, bosonization, and critical parafermionic
  theories}},
\newblock JHEP \textbf{04}, 285 (2021),
\newblock \doi{10.1007/JHEP04(2021)285},
\newblock \eprint{2012.07529}.

\bibitem{Ryu_2012}
S.~Ryu and S.-C. Zhang,
\newblock \emph{Interacting topological phases and modular invariance},
\newblock Physical Review B \textbf{85}(24) (2012),
\newblock \doi{10.1103/physrevb.85.245132}.

\bibitem{Fradkin:1980th}
E.~H. Fradkin and L.~P. Kadanoff,
\newblock \emph{{DISORDER VARIABLES AND PARAFERMIONS IN TWO-DIMENSIONAL
  STATISTICAL MECHANICS}},
\newblock Nucl. Phys. B \textbf{170}, 1 (1980),
\newblock \doi{10.1016/0550-3213(80)90472-1}.

\bibitem{Mong_2014}
R.~S. Mong, D.~J. Clarke, J.~Alicea, N.~H. Lindner, P.~Fendley, C.~Nayak,
  Y.~Oreg, A.~Stern, E.~Berg, K.~Shtengel and M.~P. Fisher,
\newblock \emph{Universal topological quantum computation from a
  superconductor-abelian quantum hall heterostructure},
\newblock Physical Review X \textbf{4}(1) (2014),
\newblock \doi{10.1103/physrevx.4.011036}.

\bibitem{Mong:2014ova}
R.~S.~K. Mong, D.~J. Clarke, J.~Alicea, N.~H. Lindner and P.~Fendley,
\newblock \emph{{Parafermionic conformal field theory on the lattice}},
\newblock J. Phys. A \textbf{47}(45), 452001 (2014),
\newblock \doi{10.1088/1751-8113/47/45/452001},
\newblock \eprint{1406.0846}.

\bibitem{Kawabata:2024hzx}
K.~Kawabata,
\newblock \emph{{Fermionic and parafermionic CFTs with $\widehat{su}(2)$ and
  $\widehat{su}(3)$ symmetry}}  (2024),
\newblock \eprint{2411.01926}.

\bibitem{Northe:2024tnm}
C.~Northe,
\newblock \emph{{Young Researchers School 2024 Maynooth: Lectures on CFT, BCFT
  and DCFT}}  (2024),
\newblock \eprint{2411.03381}.

\bibitem{Fuchs:1997af}
J.~Fuchs,
\newblock \emph{{Lectures on conformal field theory and Kac-Moody algebras}},
\newblock Lect. Notes Phys. \textbf{498}, 1 (1997),
\newblock \doi{10.1007/BFb0105277},
\newblock \eprint{hep-th/9702194}.

\bibitem{Schweigert:2000ix}
C.~Schweigert, J.~Fuchs and J.~Walcher,
\newblock \emph{{Conformal field theory, boundary conditions and applications
  to string theory}},
\newblock In \emph{{Eotvos Summer School in Physics: Nonperturbative QFT
  Methods and Their Applications}}, pp. 37--93,
\newblock \doi{10.1142/9789812799968_0002} (2000), \eprint{hep-th/0011109}.

\bibitem{Bais:2008ni}
F.~A. Bais and J.~K. Slingerland,
\newblock \emph{{Condensate induced transitions between topologically ordered
  phases}},
\newblock Phys. Rev. B \textbf{79}, 045316 (2009),
\newblock \doi{10.1103/PhysRevB.79.045316},
\newblock \eprint{0808.0627}.

\bibitem{2013PhRvX...3b1009L}
M.~{Levin},
\newblock \emph{{Protected Edge Modes without Symmetry}},
\newblock Physical Review X \textbf{3}(2), 021009 (2013),
\newblock \doi{10.1103/PhysRevX.3.021009},
\newblock \eprint{1301.7355}.

\bibitem{Kong:2013aya}
L.~Kong,
\newblock \emph{{Anyon condensation and tensor categories}},
\newblock Nucl. Phys. B \textbf{886}, 436 (2014),
\newblock \doi{10.1016/j.nuclphysb.2014.07.003},
\newblock \eprint{1307.8244}.

\bibitem{Fuchs:1993et}
J.~Fuchs,
\newblock \emph{{Fusion rules in conformal field theory}},
\newblock Fortsch. Phys. \textbf{42}, 1 (1994),
\newblock \doi{10.1002/prop.2190420102},
\newblock \eprint{hep-th/9306162}.

\bibitem{Rida:1999xu}
A.~Rida and T.~Sami,
\newblock \emph{{The nonchiral fusion rules in rational conformal field
  theories}},
\newblock Lett. Math. Phys. \textbf{58}, 239 (2001),
\newblock \doi{10.1023/A:1014599117130},
\newblock \eprint{hep-th/9907137}.

\bibitem{Rida:1999ru}
A.~Rida and T.~Sami,
\newblock \emph{{Nonchiral fusion rules, structure constants of D(m) minimal
  models}}  (1999),
\newblock \eprint{hep-th/9910070}.

\bibitem{Moore:1989vd}
G.~W. Moore and N.~Seiberg,
\newblock \emph{{Lectures on RCFT}},
\newblock In \emph{{Strings '89, Proceedings of the Trieste Spring School on
  Superstrings.}} World Scientific (1990).

\bibitem{Polyakov:1974gs}
A.~M. Polyakov,
\newblock \emph{{Nonhamiltonian approach to conformal quantum field theory}},
\newblock Zh. Eksp. Teor. Fiz. \textbf{66}, 23 (1974).

\bibitem{Nivesvivat:2025odb}
R.~Nivesvivat and S.~Ribault,
\newblock \emph{{Fusion rules and structure constants of E-series minimal
  models}},
\newblock SciPost Phys. \textbf{18}, 163 (2025),
\newblock \doi{10.21468/SciPostPhys.18.5.163},
\newblock \eprint{2502.14295}.

\bibitem{Huang:2005gz}
Y.-Z. Huang and L.~Kong,
\newblock \emph{{Full field algebras}},
\newblock Commun. Math. Phys. \textbf{272}, 345 (2007),
\newblock \doi{10.1007/s00220-007-0224-4},
\newblock \eprint{math/0511328}.

\bibitem{Kong:2006wa}
L.~Kong,
\newblock \emph{{Full field algebras, operads and tensor categories}},
\newblock Adv. Math. \textbf{213}, 271 (2007),
\newblock \doi{10.1016/j.aim.2006.12.007},
\newblock \eprint{math/0603065}.

\bibitem{Huang:2006ar}
Y.-Z. Huang and L.~Kong,
\newblock \emph{{Modular invariance for conformal full field algebras}},
\newblock Trans. Amer. Math. Soc. \textbf{362}, 3027 (2010),
\newblock \doi{10.1090/S0002-9947-09-04933-2},
\newblock \eprint{math/0609570}.

\bibitem{Moriwaki_2023}
Y.~Moriwaki,
\newblock \emph{Two-dimensional conformal field theory, full vertex algebra and
  current-current deformation},
\newblock Advances in Mathematics \textbf{427}, 109125 (2023),
\newblock \doi{10.1016/j.aim.2023.109125}.

\bibitem{Vicedo:2025vql}
B.~Vicedo,
\newblock \emph{{Full affine Kac-Moody vertex algebra from factorisation}}
  (2025),
\newblock \eprint{2501.08412}.

\bibitem{Verlinde:1988sn}
E.~P. Verlinde,
\newblock \emph{{Fusion Rules and Modular Transformations in 2D Conformal Field
  Theory}},
\newblock Nucl. Phys. B \textbf{300}, 360 (1988),
\newblock \doi{10.1016/0550-3213(88)90603-7}.

\bibitem{Yao:2019bub}
Y.~Yao and Y.~Fukusumi,
\newblock \emph{{Bosonization with a background $U(1)$ gauge field}},
\newblock Phys. Rev. B \textbf{100}(7), 075105 (2019),
\newblock \doi{10.1103/PhysRevB.100.075105},
\newblock \eprint{1902.06584}.

\bibitem{Fukusumi:2020irh}
Y.~Fukusumi and S.~Iino,
\newblock \emph{{Open spin chain realization of a topological defect in a
  one-dimensional Ising model: Boundary and bulk symmetry}},
\newblock Phys. Rev. B \textbf{104}(12), 125418 (2021),
\newblock \doi{10.1103/PhysRevB.104.125418},
\newblock \eprint{2004.04415}.

\bibitem{Kapustin:2017jrc}
A.~Kapustin and R.~Thorngren,
\newblock \emph{{Fermionic SPT phases in higher dimensions and bosonization}},
\newblock JHEP \textbf{10}, 080 (2017),
\newblock \doi{10.1007/JHEP10(2017)080},
\newblock \eprint{1701.08264}.

\bibitem{Karch:2019lnn}
A.~Karch, D.~Tong and C.~Turner,
\newblock \emph{{A Web of 2d Dualities: ${\bf Z}_2$ Gauge Fields and Arf
  Invariants}},
\newblock SciPost Phys. \textbf{7}, 007 (2019),
\newblock \doi{10.21468/SciPostPhys.7.1.007},
\newblock \eprint{1902.05550}.

\bibitem{Apruzzi:2021nmk}
F.~Apruzzi, F.~Bonetti, I.~Garc{\'\i}a~Etxebarria, S.~S. Hosseini and
  S.~Schafer-Nameki,
\newblock \emph{{Symmetry TFTs from String Theory}},
\newblock Commun. Math. Phys. \textbf{402}(1), 895 (2023),
\newblock \doi{10.1007/s00220-023-04737-2},
\newblock \eprint{2112.02092}.

\bibitem{Barkeshli:2014cna}
M.~Barkeshli, P.~Bonderson, M.~Cheng and Z.~Wang,
\newblock \emph{{Symmetry Fractionalization, Defects, and Gauging of
  Topological Phases}},
\newblock Phys. Rev. B \textbf{100}(11), 115147 (2019),
\newblock \doi{10.1103/PhysRevB.100.115147},
\newblock \eprint{1410.4540}.

\bibitem{davydov2021braidedpicardgroupsgraded}
A.~Davydov and D.~Nikshych,
\newblock \emph{Braided picard groups and graded extensions of braided tensor
  categories} (2021), \eprint{2006.08022}.

\bibitem{Bischoff:2019jho}
M.~Bischoff and C.~Jones,
\newblock \emph{{Computing fusion rules for spherical G-extensions of fusion
  categories}},
\newblock Selecta Math. \textbf{28}(2), 26 (2022),
\newblock \doi{10.1007/s00029-021-00725-3},
\newblock \eprint{1909.02816}.

\bibitem{Carqueville:2025kqs}
N.~Carqueville and B.~Haake,
\newblock \emph{{2-Group Symmetries of 3-dimensional Defect TQFTs and Their
  Gauging}}  (2025),
\newblock \eprint{2506.08178}.

\bibitem{Heinrich:2025wkx}
S.~Heinrich, J.~Plavnik, I.~Runkel and A.~Watkins,
\newblock \emph{{Generalised Orbifolds and G-equivariantisation}}  (2025),
\newblock \eprint{2506.08154}.

\bibitem{Chang:2018iay}
C.-M. Chang, Y.-H. Lin, S.-H. Shao, Y.~Wang and X.~Yin,
\newblock \emph{{Topological Defect Lines and Renormalization Group Flows in
  Two Dimensions}},
\newblock JHEP \textbf{01}, 026 (2019),
\newblock \doi{10.1007/JHEP01(2019)026},
\newblock \eprint{1802.04445}.

\bibitem{Komargodski:2020mxz}
Z.~Komargodski, K.~Ohmori, K.~Roumpedakis and S.~Seifnashri,
\newblock \emph{{Symmetries and strings of adjoint QCD$_{2}$}},
\newblock JHEP \textbf{03}, 103 (2021),
\newblock \doi{10.1007/JHEP03(2021)103},
\newblock \eprint{2008.07567}.

\bibitem{Wess:1971yu}
J.~Wess and B.~Zumino,
\newblock \emph{{Consequences of anomalous Ward identities}},
\newblock Phys. Lett. B \textbf{37}, 95 (1971),
\newblock \doi{10.1016/0370-2693(71)90582-X}.

\bibitem{Witten:1983tw}
E.~Witten,
\newblock \emph{{Global Aspects of Current Algebra}},
\newblock Nucl. Phys. B \textbf{223}, 422 (1983),
\newblock \doi{10.1016/0550-3213(83)90063-9}.

\bibitem{Witten:1983ar}
E.~Witten,
\newblock \emph{{Nonabelian Bosonization in Two-Dimensions}},
\newblock Commun. Math. Phys. \textbf{92}, 455 (1984),
\newblock \doi{10.1007/BF01215276}.

\bibitem{Furuya:2015coa}
S.~C. Furuya and M.~Oshikawa,
\newblock \emph{{Symmetry Protection of Critical Phases and a Global Anomaly in
  $1+1$ Dimensions}},
\newblock Phys. Rev. Lett. \textbf{118}(2), 021601 (2017),
\newblock \doi{10.1103/PhysRevLett.118.021601},
\newblock \eprint{1503.07292}.

\bibitem{Numasawa:2017crf}
T.~Numasawa and S.~Yamaguch,
\newblock \emph{{Mixed Global Anomalies and Boundary Conformal Field
  Theories}},
\newblock JHEP \textbf{11}, 202 (2018),
\newblock \doi{10.1007/JHEP11(2018)202},
\newblock \eprint{1712.09361}.

\bibitem{Kikuchi:2019ytf}
K.~Kikuchi and Y.~Zhou,
\newblock \emph{{Two-dimensional Anomaly, Orbifolding, and Boundary States}}
  (2019),
\newblock \eprint{1908.02918}.

\bibitem{Fateev:1985mm}
V.~A. Fateev and A.~B. Zamolodchikov,
\newblock \emph{{Parafermionic Currents in the Two-Dimensional Conformal
  Quantum Field Theory and Selfdual Critical Points in Z(n) Invariant
  Statistical Systems}},
\newblock Sov. Phys. JETP \textbf{62}, 215 (1985).

\bibitem{Lu_2010}
Y.-M. Lu, X.-G. Wen, Z.~Wang and Z.~Wang,
\newblock \emph{Non-abelian quantum hall states and their quasiparticles: From
  the pattern of zeros to vertex algebra},
\newblock Physical Review B \textbf{81}(11) (2010),
\newblock \doi{10.1103/physrevb.81.115124}.

\bibitem{Gannon:2003de}
T.~Gannon,
\newblock \emph{{Comments on nonunitary conformal field theories}},
\newblock Nucl. Phys. B \textbf{670}, 335 (2003),
\newblock \doi{10.1016/j.nuclphysb.2003.07.030},
\newblock \eprint{hep-th/0305070}.

\bibitem{Duan:2023ykn}
Z.~Duan, Q.~Jia and S.~Lee,
\newblock \emph{{{\ensuremath{\mathbb{Z}}}$_{N}$ duality and parafermions
  revisited}},
\newblock JHEP \textbf{11}, 206 (2023),
\newblock \doi{10.1007/JHEP11(2023)206},
\newblock \eprint{2309.01913}.

\bibitem{Chen:2023jht}
J.~Chen, B.~Haghighat and Q.-R. Wang,
\newblock \emph{{Para-fusion Category and Topological Defect Lines in $\mathbb
  Z_N$-parafermionic CFTs}}  (2023),
\newblock \eprint{2309.01914}.

\bibitem{Huang:2023pyk}
S.-J. Huang and M.~Cheng,
\newblock \emph{{Topological holography, quantum criticality, and boundary
  states}}  (2023),
\newblock \eprint{2310.16878}.

\bibitem{Bhardwaj:2024ydc}
L.~Bhardwaj, K.~Inamura and A.~Tiwari,
\newblock \emph{{Fermionic Non-Invertible Symmetries in (1+1)d: Gapped and
  Gapless Phases, Transitions, and Symmetry TFTs}}  (2024),
\newblock \eprint{2405.09754}.

\bibitem{Huang:2024ror}
S.-J. Huang,
\newblock \emph{{Fermionic quantum criticality through the lens of topological
  holography}}  (2024),
\newblock \eprint{2405.09611}.

\bibitem{KNBalasubramanian:2025vpe}
M.~K.~N.~Balasubramanian, M.~Buican, C.~Delcamp and R.~Radhakrishnan,
\newblock \emph{{Gauging Non-Invertible Symmetries in (2+1)d Topological
  Orders}}  (2025),
\newblock \eprint{2507.01142}.

\bibitem{Moore:1988qv}
G.~W. Moore and N.~Seiberg,
\newblock \emph{{Classical and Quantum Conformal Field Theory}},
\newblock Commun. Math. Phys. \textbf{123}, 177 (1989),
\newblock \doi{10.1007/BF01238857}.

\bibitem{Moore:1988ss}
G.~W. Moore and N.~Seiberg,
\newblock \emph{{Naturality in Conformal Field Theory}},
\newblock Nucl. Phys. B \textbf{313}, 16 (1989),
\newblock \doi{10.1016/0550-3213(89)90511-7}.

\bibitem{Chatterjee:2022kxb}
A.~Chatterjee and X.-G. Wen,
\newblock \emph{{Symmetry as a shadow of topological order and a derivation of
  topological holographic principle}},
\newblock Phys. Rev. B \textbf{107}(15), 155136 (2023),
\newblock \doi{10.1103/PhysRevB.107.155136},
\newblock \eprint{2203.03596}.

\bibitem{Kong:2022cpy}
L.~Kong and Z.-H. Zhang,
\newblock \emph{{An invitation to topological orders and category theory}}
  (2022),
\newblock \eprint{2205.05565}.

\bibitem{Kong:2014qka}
L.~Kong and X.-G. Wen,
\newblock \emph{{Braided fusion categories, gravitational anomalies, and the
  mathematical framework for topological orders in any dimensions}}  (2014),
\newblock \eprint{1405.5858}.

\bibitem{kong2015boundarybulkrelationtopologicalorders}
L.~Kong, X.-G. Wen and H.~Zheng,
\newblock \emph{Boundary-bulk relation for topological orders as the functor
  mapping higher categories to their centers} (2015), \eprint{1502.01690}.

\bibitem{Kong:2017etd}
L.~Kong and H.~Zheng,
\newblock \emph{{Gapless edges of 2d topological orders and enriched monoidal
  categories}},
\newblock Nucl. Phys. B \textbf{927}, 140 (2018),
\newblock \doi{10.1016/j.nuclphysb.2017.12.007},
\newblock \eprint{1705.01087}.

\bibitem{Kong:2017hcw}
L.~Kong, X.-G. Wen and H.~Zheng,
\newblock \emph{{Boundary-bulk relation in topological orders}},
\newblock Nucl. Phys. B \textbf{922}, 62 (2017),
\newblock \doi{10.1016/j.nuclphysb.2017.06.023},
\newblock \eprint{1702.00673}.

\bibitem{Cardy:2017ufe}
J.~Cardy,
\newblock \emph{{Bulk Renormalization Group Flows and Boundary States in
  Conformal Field Theories}},
\newblock SciPost Phys. \textbf{3}(2), 011 (2017),
\newblock \doi{10.21468/SciPostPhys.3.2.011},
\newblock \eprint{1706.01568}.

\bibitem{Date:1987zz}
E.~Date, M.~Jimbo, T.~Miwa and M.~Okado,
\newblock \emph{{Automorphic properties of local height probabilities for
  integrable solid-on-solid models}},
\newblock Phys. Rev. B \textbf{35}, 2105 (1987),
\newblock \doi{10.1103/PhysRevB.35.2105}.

\bibitem{Saleur:1988zx}
H.~Saleur and M.~Bauer,
\newblock \emph{{On Some Relations Between Local Height Probabilities and
  Conformal Invariance}},
\newblock Nucl. Phys. B \textbf{320}, 591 (1989),
\newblock \doi{10.1016/0550-3213(89)90014-X}.

\bibitem{Foda:2017vog}
O.~Foda,
\newblock \emph{{Off-critical local height probabilities on a plane and
  critical partition functions on a cylinder}}  (2017),
\newblock \doi{10.1016/j.nuclphysb.2018.01.011},
\newblock \eprint{1711.03337}.

\bibitem{Cardy:1986gw}
J.~L. Cardy,
\newblock \emph{{Effect of Boundary Conditions on the Operator Content of
  Two-Dimensional Conformally Invariant Theories}},
\newblock Nucl. Phys. B \textbf{275}, 200 (1986),
\newblock \doi{10.1016/0550-3213(86)90596-1}.

\bibitem{Cardy:1989ir}
J.~L. Cardy,
\newblock \emph{{Boundary Conditions, Fusion Rules and the Verlinde Formula}},
\newblock Nucl. Phys. B \textbf{324}, 581 (1989),
\newblock \doi{10.1016/0550-3213(89)90521-X}.

\bibitem{Lencses:2018paa}
M.~Lencses, J.~Viti and G.~Takacs,
\newblock \emph{{Chiral entanglement in massive quantum field theories in 1+1
  dimensions}},
\newblock JHEP \textbf{01}, 177 (2019),
\newblock \doi{10.1007/JHEP01(2019)177},
\newblock \eprint{1811.06500}.

\bibitem{Ares:2020uwy}
F.~Ares, M.~A. Rajabpour and J.~Viti,
\newblock \emph{{Scaling of the Formation Probabilities and Universal Boundary
  Entropies in the Quantum XY Spin Chain}},
\newblock J. Stat. Mech. \textbf{2008}, 083111 (2020),
\newblock \doi{10.1088/1742-5468/aba9d4},
\newblock \eprint{2004.10606}.

\bibitem{Li:2022drc}
L.~Li, C.-T. Hsieh, Y.~Yao and M.~Oshikawa,
\newblock \emph{{Boundary conditions and anomalies of conformal field theories
  in 1+1 dimensions}}  (2022),
\newblock \eprint{2205.11190}.

\bibitem{Cordova:2022lms}
C.~Cordova and D.~Garc\'\i{}a-Sep\'ulveda,
\newblock \emph{{Symmetry Enriched $c$-Theorems \& SPT Transitions}}  (2022),
\newblock \eprint{2210.01135}.

\bibitem{Kikuchi:2021qxz}
K.~Kikuchi,
\newblock \emph{{Symmetry enhancement in RCFT}}  (2021),
\newblock \eprint{2109.02672}.

\bibitem{Kikuchi:2022biw}
K.~Kikuchi,
\newblock \emph{{Emergent symmetry and free energy}}  (2022),
\newblock \eprint{2207.10095}.

\bibitem{Kikuchi:2022gfi}
K.~Kikuchi,
\newblock \emph{{Symmetry enhancement in RCFT II}}  (2022),
\newblock \eprint{2207.06433}.

\bibitem{Kikuchi:2022ipr}
K.~Kikuchi,
\newblock \emph{{RG flows from WZW models}}  (2022),
\newblock \eprint{2212.13851}.

\bibitem{Wen:2025xka}
X.~Wen,
\newblock \emph{{Space of conformal boundary conditions from the view of higher
  Berry phase: Flow of Berry curvature in parametrized BCFTs}}  (2025),
\newblock \eprint{2507.12546}.

\bibitem{Choi:2025ebk}
Y.~Choi, H.~Ha, D.~Kim, Y.~Kusuki, S.~Ohyama and S.~Ryu,
\newblock \emph{{Higher Structures on Boundary Conformal Manifolds: Higher
  Berry Phase and Boundary Conformal Field Theory}}  (2025),
\newblock \eprint{2507.12525}.

\bibitem{Nishioka:2022ook}
T.~Nishioka, Y.~Okuyama and S.~Shimamori,
\newblock \emph{{Method of images in defect conformal field theories}},
\newblock Phys. Rev. D \textbf{106}(8), L081701 (2022),
\newblock \doi{10.1103/PhysRevD.106.L081701},
\newblock \eprint{2205.05370}.

\bibitem{Kong:2024ykr}
L.~Kong, Z.-H. Zhang, J.~Zhao and H.~Zheng,
\newblock \emph{{Higher condensation theory}}  (2024),
\newblock \eprint{2403.07813}.

\bibitem{Smith:2021luc}
P.~Boyle~Smith,
\newblock \emph{{Boundary States and Anomalous Symmetries of Fermionic Minimal
  Models}}  (2021),
\newblock \eprint{2102.02203}.

\bibitem{Weizmann}
H.~Ebisu and M.~Watanabe,
\newblock \emph{{Fermionization of conformal boundary states}},
\newblock Phys. Rev. B \textbf{104}(19), 195124 (2021),
\newblock \doi{10.1103/PhysRevB.104.195124},
\newblock \eprint{2103.01101}.

\bibitem{Fukusumi:2021zme}
Y.~Fukusumi, Y.~Tachikawa and Y.~Zheng,
\newblock \emph{{Fermionization and boundary states in 1+1 dimensions}},
\newblock SciPost Phys. \textbf{11}, 082 (2021),
\newblock \doi{10.21468/SciPostPhys.11.4.082},
\newblock \eprint{2103.00746}.

\bibitem{Deligne1990}
P.~Deligne,
\newblock \emph{{Cat\'{e}gories tannakiennes, The Grothendieck Festschrift
  volume 2}},
\newblock Progr. Math. \textbf{87}, 111 (1990).

\bibitem{PhysRevLett.54.1091}
L.-F. Ko, H.~Au-Yang and J.~H.~H. Perk,
\newblock \emph{Energy-density correlation functions in the two-dimensional
  ising model with a line defect},
\newblock Phys. Rev. Lett. \textbf{54}, 1091 (1985),
\newblock \doi{10.1103/PhysRevLett.54.1091}.

\bibitem{Stern2003GeometricPA}
A.~Stern, F.~von Oppen and E.~Mariani,
\newblock \emph{Geometric phases and quantum entanglement as building blocks
  for non-abelian quasiparticle statistics},
\newblock Physical Review B \textbf{70}, 205338 (2003).

\bibitem{Schellekens:1996tg}
A.~N. Schellekens,
\newblock \emph{{Introduction to conformal field theory}},
\newblock Fortsch. Phys. \textbf{44}, 605 (1996).

\bibitem{Lou:2020gfq}
J.~Lou, C.~Shen, C.~Chen and L.-Y. Hung,
\newblock \emph{{A (dummy\textquoteright{}s) guide to working with gapped
  boundaries via (fermion) condensation}},
\newblock JHEP \textbf{02}, 171 (2021),
\newblock \doi{10.1007/JHEP02(2021)171},
\newblock \eprint{2007.10562}.

\bibitem{Kong:2019cuu}
L.~Kong and H.~Zheng,
\newblock \emph{{A mathematical theory of gapless edges of 2d topological
  orders. Part II}},
\newblock Nucl. Phys. B \textbf{966}, 115384 (2021),
\newblock \doi{10.1016/j.nuclphysb.2021.115384},
\newblock \eprint{1912.01760}.

\bibitem{Barkeshli:2015afa}
M.~Barkeshli, M.~Mulligan and M.~P.~A. Fisher,
\newblock \emph{{Particle-Hole Symmetry and the Composite Fermi Liquid}},
\newblock Phys. Rev. B \textbf{92}(16), 165125 (2015),
\newblock \doi{10.1103/PhysRevB.92.165125},
\newblock \eprint{1502.05404}.

\bibitem{Son:2015xqa}
D.~T. Son,
\newblock \emph{{Is the Composite Fermion a Dirac Particle?}},
\newblock Phys. Rev. X \textbf{5}(3), 031027 (2015),
\newblock \doi{10.1103/PhysRevX.5.031027},
\newblock \eprint{1502.03446}.

\bibitem{Wang2017TopologicalOF}
C.~Wang, A.~Vishwanath and B.~I. Halperin,
\newblock \emph{Topological order from disorder and the quantized hall thermal
  metal: Possible applications to the $\nu$=5/2 state},
\newblock Physical Review B  (2017).

\bibitem{Mross_2018}
D.~F. Mross, Y.~Oreg, A.~Stern, G.~Margalit and M.~Heiblum,
\newblock \emph{Theory of disorder-induced half-integer thermal hall
  conductance},
\newblock Physical Review Letters \textbf{121}(2) (2018),
\newblock \doi{10.1103/physrevlett.121.026801}.

\bibitem{Stone:2012ud}
M.~Stone,
\newblock \emph{{Gravitational Anomalies and Thermal Hall effect in Topological
  Insulators}},
\newblock Phys. Rev. B \textbf{85}, 184503 (2012),
\newblock \doi{10.1103/PhysRevB.85.184503},
\newblock \eprint{1201.4095}.

\bibitem{Callan:1984sa}
C.~G. Callan, Jr. and J.~A. Harvey,
\newblock \emph{{Anomalies and Fermion Zero Modes on Strings and Domain
  Walls}},
\newblock Nucl. Phys. B \textbf{250}, 427 (1985),
\newblock \doi{10.1016/0550-3213(85)90489-4}.

\bibitem{Kaidi:2021gbs}
J.~Kaidi, Z.~Komargodski, K.~Ohmori, S.~Seifnashri and S.-H. Shao,
\newblock \emph{{Higher central charges and topological boundaries in
  2+1-dimensional TQFTs}}  (2021),
\newblock \eprint{2107.13091}.

\bibitem{Wong:1994np}
E.~Wong and I.~Affleck,
\newblock \emph{{Tunneling in quantum wires: A Boundary conformal field theory
  approach}},
\newblock Nucl. Phys. B \textbf{417}, 403 (1994),
\newblock \doi{10.1016/0550-3213(94)90479-0},
\newblock \eprint{cond-mat/9311040}.

\bibitem{Fuchs:1999xn}
J.~Fuchs and C.~Schweigert,
\newblock \emph{{Symmetry breaking boundaries. 2. More structures: Examples}},
\newblock Nucl. Phys. B \textbf{568}, 543 (2000),
\newblock \doi{10.1016/S0550-3213(99)00669-0},
\newblock \eprint{hep-th/9908025}.

\bibitem{Ishikawa:2002wx}
H.~Ishikawa and T.~Tani,
\newblock \emph{{Novel construction of boundary states in coset conformal field
  theories}},
\newblock Nucl. Phys. B \textbf{649}, 205 (2003),
\newblock \doi{10.1016/S0550-3213(02)01011-8},
\newblock \eprint{hep-th/0207177}.

\bibitem{Ishikawa:2005ea}
H.~Ishikawa and T.~Tani,
\newblock \emph{{Twisted Boundary States and Representation of Generalized
  Fusion Algebra}},
\newblock Nucl. Phys. B \textbf{739}, 328 (2006),
\newblock \doi{10.1016/j.nuclphysb.2006.01.031},
\newblock \eprint{hep-th/0510242}.

\bibitem{Behrend:2000us}
R.~E. Behrend and P.~A. Pearce,
\newblock \emph{{Integrable and conformal boundary conditions for sl(2) A-D-E
  lattice models and unitary minimal conformal field theories}},
\newblock J. Statist. Phys. \textbf{102}, 577 (2001),
\newblock \doi{10.1023/A:1004890600991},
\newblock \eprint{hep-th/0006094}.

\bibitem{Iino:2020ipa}
S.~Iino,
\newblock \emph{{Boundary CFT and tensor network approach to surface critical
  phenomena of the tricritical 3-state Potts model}},
\newblock J. Statist. Phys. \textbf{182}(3), 56 (2021),
\newblock \doi{10.1007/s10955-021-02728-y},
\newblock \eprint{2007.03182}.

\bibitem{Recknagel:2013uja}
A.~Recknagel and V.~Schomerus,
\newblock \emph{{Boundary Conformal Field Theory and the Worldsheet Approach to
  D-Branes}},
\newblock Cambridge Monographs on Mathematical Physics. Cambridge University
  Press,
\newblock ISBN 978-0-521-83223-6, 978-0-521-83223-6, 978-1-107-49612-5,
\newblock \doi{10.1017/CBO9780511806476} (2013).

\bibitem{Quella:2006de}
T.~Quella, I.~Runkel and G.~M.~T. Watts,
\newblock \emph{{Reflection and transmission for conformal defects}},
\newblock JHEP \textbf{04}, 095 (2007),
\newblock \doi{10.1088/1126-6708/2007/04/095},
\newblock \eprint{hep-th/0611296}.

\bibitem{Kimura:2014hva}
T.~Kimura and M.~Murata,
\newblock \emph{{Current Reflection and Transmission at Conformal Defects:
  Applying BCFT to Transport Process}},
\newblock Nucl. Phys. B \textbf{885}, 266 (2014),
\newblock \doi{10.1016/j.nuclphysb.2014.05.026},
\newblock \eprint{1402.6705}.

\bibitem{Kimura:2015nka}
T.~Kimura and M.~Murata,
\newblock \emph{{Transport Process in Multi-Junctions of Quantum Systems}},
\newblock JHEP \textbf{07}, 072 (2015),
\newblock \doi{10.1007/JHEP07(2015)072},
\newblock \eprint{1505.05275}.

\bibitem{Stanishkov:2016pvi}
M.~Stanishkov,
\newblock \emph{{RG domain wall for the general $ \widehat{su}(2) $ coset
  models}},
\newblock JHEP \textbf{08}, 096 (2016),
\newblock \doi{10.1007/JHEP08(2016)096},
\newblock \eprint{1606.03605}.

\bibitem{Poghosyan:2022mfw}
H.~Poghosyan and R.~Poghossian,
\newblock \emph{{RG flow between W$_{3}$ minimal models by perturbation and
  domain wall approaches}},
\newblock JHEP \textbf{08}, 307 (2022),
\newblock \doi{10.1007/JHEP08(2022)307},
\newblock \eprint{2205.05091}.

\bibitem{Poghosyan:2023brb}
A.~Poghosyan and H.~Poghosyan,
\newblock \emph{{A note on RG domain wall between successive $ {A}_2^{(p)} $
  minimal models}},
\newblock JHEP \textbf{08}, 072 (2023),
\newblock \doi{10.1007/JHEP08(2023)072},
\newblock \eprint{2305.05997}.

\bibitem{Lan:2014uaa}
T.~Lan, J.~C. Wang and X.-G. Wen,
\newblock \emph{{Gapped Domain Walls, Gapped Boundaries and Topological
  Degeneracy}},
\newblock Phys. Rev. Lett. \textbf{114}(7), 076402 (2015),
\newblock \doi{10.1103/PhysRevLett.114.076402},
\newblock \eprint{1408.6514}.

\bibitem{Wan:2016php}
Y.~Wan and C.~Wang,
\newblock \emph{{Fermion Condensation and Gapped Domain Walls in Topological
  Orders}},
\newblock JHEP \textbf{03}, 172 (2017),
\newblock \doi{10.1007/JHEP03(2017)172},
\newblock \eprint{1607.01388}.

\bibitem{Zhao:2023wtg}
Y.~Zhao, H.~Wang, Y.~Hu and Y.~Wan,
\newblock \emph{{Symmetry fractionalized (irrationalized) fusion rules and two
  domain-wall Verlinde formulae}},
\newblock JHEP \textbf{04}, 115 (2024),
\newblock \doi{10.1007/JHEP04(2024)115},
\newblock \eprint{2304.08475}.

\bibitem{Jia:2025yph}
Z.~Jia and S.~Tan,
\newblock \emph{{Weak Hopf tube algebra for domain walls between 2d gapped
  phases of Turaev-Viro TQFTs}}  (2025),
\newblock \eprint{2507.01515}.

\bibitem{Buican:2025zpm}
M.~Buican, R.~Geiko, M.~Moses and B.~Shi,
\newblock \emph{{An Algebraic Theory of Gapped Domain Wall Partons}}  (2025),
\newblock \eprint{2506.22544}.

\bibitem{Cordova:2025eim}
C.~Cordova, D.~Garc\'\i{}a-Sep\'ulveda and K.~Ohmori,
\newblock \emph{{Higgsing Transitions from Topological Field Theory \&
  Non-Invertible Symmetry in Chern-Simons Matter Theories}}  (2025),
\newblock \eprint{2504.03614}.

\bibitem{Antinucci:2025uvj}
A.~Antinucci, C.~Copetti, G.~Galati and G.~Rizi,
\newblock \emph{{Defect Conformal Manifolds from Phantom (Non-Invertible)
  Symmetries}}  (2025),
\newblock \eprint{2505.09668}.

\bibitem{Antinucci:2025fjp}
A.~Antinucci, C.~Copetti, Y.~Gai and S.~Schafer-Nameki,
\newblock \emph{{Categorical Anomaly Matching}}  (2025),
\newblock \eprint{2508.00982}.

\bibitem{KNBalasubramanian:2025vum}
M.~K.~N.~Balasubramanian, M.~Buican, C.~Delcamp and R.~Radhakrishnan,
\newblock \emph{{Gauging Non-Invertible Symmetries in (2+1)d Topological
  Orders}}  (2025),
\newblock \eprint{2507.01142}.

\bibitem{Nakayama:2024msv}
Y.~Nakayama and T.~Tanaka,
\newblock \emph{{Infinitely many new renormalization group flows between
  Virasoro minimal models from non-invertible symmetries}},
\newblock JHEP \textbf{11}, 137 (2024),
\newblock \doi{10.1007/JHEP11(2024)137},
\newblock \eprint{2407.21353}.

\bibitem{Lu:2022ver}
D.-C. Lu and Z.~Sun,
\newblock \emph{{On triality defects in 2d CFT}},
\newblock JHEP \textbf{02}, 173 (2023),
\newblock \doi{10.1007/JHEP02(2023)173},
\newblock \eprint{2208.06077}.

\bibitem{Perez-Lona:2023djo}
A.~Perez-Lona, D.~Robbins, E.~Sharpe, T.~Vandermeulen and X.~Yu,
\newblock \emph{{Notes on gauging noninvertible symmetries. Part I.
  Multiplicity-free cases}},
\newblock JHEP \textbf{02}, 154 (2024),
\newblock \doi{10.1007/JHEP02(2024)154},
\newblock \eprint{2311.16230}.

\bibitem{Choi:2023vgk}
Y.~Choi, D.-C. Lu and Z.~Sun,
\newblock \emph{{Self-duality under gauging a non-invertible symmetry}},
\newblock JHEP \textbf{01}, 142 (2024),
\newblock \doi{10.1007/JHEP01(2024)142},
\newblock \eprint{2310.19867}.

\bibitem{Diatlyk:2023fwf}
O.~Diatlyk, C.~Luo, Y.~Wang and Q.~Weller,
\newblock \emph{{Gauging non-invertible symmetries: topological interfaces and
  generalized orbifold groupoid in 2d QFT}},
\newblock JHEP \textbf{03}, 127 (2024),
\newblock \doi{10.1007/JHEP03(2024)127},
\newblock \eprint{2311.17044}.

\bibitem{Perez-Lona:2024sds}
A.~Perez-Lona, D.~Robbins, E.~Sharpe, T.~Vandermeulen and X.~Yu,
\newblock \emph{{Notes on gauging noninvertible symmetries. Part II. Higher
  multiplicity cases}},
\newblock JHEP \textbf{05}, 066 (2025),
\newblock \doi{10.1007/JHEP05(2025)066},
\newblock \eprint{2408.16811}.

\bibitem{Lu:2024ytl}
D.-C. Lu, Z.~Sun and Y.-Z. You,
\newblock \emph{{Realizing triality and $p$-ality by lattice twisted gauging in
  (1+1)d quantum spin systems}},
\newblock SciPost Phys. \textbf{17}(5), 136 (2024),
\newblock \doi{10.21468/SciPostPhys.17.5.136},
\newblock \eprint{2405.14939}.

\bibitem{Ando:2024hun}
T.~Ando,
\newblock \emph{{A journey on self-$G$-ality}}  (2024),
\newblock \eprint{2405.15648}.

\bibitem{Lu:2024lzf}
D.-C. Lu, Z.~Sun and Z.~Zhang,
\newblock \emph{{Exploring $G$-ality defects in 2-dim QFTs}}  (2024),
\newblock \eprint{2406.12151}.

\bibitem{Yu:2025iqf}
X.~Yu and H.~Y. Zhang,
\newblock \emph{{von Neumann Subfactors and Non-invertible Symmetries}}
  (2025),
\newblock \eprint{2504.05374}.

\bibitem{Maeda:2025rxc}
J.~Maeda and T.~Oishi,
\newblock \emph{{$N$-ality symmetry and SPT phases in (1+1)d}}  (2025),
\newblock \eprint{2504.20151}.

\bibitem{Lu:2025gpt}
D.-C. Lu, Z.~Sun and Z.~Zhang,
\newblock \emph{{SymSETs and self-dualities under gauging non-invertible
  symmetries}}  (2025),
\newblock \eprint{2501.07787}.

\bibitem{Seifnashri:2025fgd}
S.~Seifnashri, S.-H. Shao and X.~Yang,
\newblock \emph{{Gauging non-invertible symmetries on the lattice}}  (2025),
\newblock \eprint{2503.02925}.

\bibitem{Tanaka:2025qou}
T.~Tanaka and Y.~Nakayama,
\newblock \emph{{ADE triality via (non-)invertible symmetry gauging}}  (2025),
\newblock \eprint{2506.15158}.

\bibitem{Galindo:2024qzg}
C.~Galindo, S.~Lentner and S.~M\"oller,
\newblock \emph{{Computing $G$-Crossed Extensions and Orbifolds of Vertex
  Operator Algebras}}  (2024),
\newblock \eprint{2409.16357}.

\bibitem{Haldane:1981zza}
F.~D.~M. Haldane,
\newblock \emph{{Luttinger liquid theory of one-dimensional quantum fluids. I.
  Properties of the Luttinger model and their extension to the general 1D
  interacting spinless Fermi gas}},
\newblock J. Phys. C \textbf{14}, 2585 (1981),
\newblock \doi{10.1088/0022-3719/14/19/010}.

\bibitem{Haldane:1982rj}
F.~D.~M. Haldane,
\newblock \emph{{Continuum dynamics of the 1-D Heisenberg antiferromagnetic
  identification with the O(3) nonlinear sigma model}},
\newblock Phys. Lett. A \textbf{93}, 464 (1983),
\newblock \doi{10.1016/0375-9601(83)90631-X}.

\bibitem{Haldane:1983ru}
F.~D.~M. Haldane,
\newblock \emph{{Nonlinear field theory of large spin Heisenberg
  antiferromagnets. Semiclassically quantized solitons of the one-dimensional
  easy Axis Neel state}},
\newblock Phys. Rev. Lett. \textbf{50}, 1153 (1983),
\newblock \doi{10.1103/PhysRevLett.50.1153}.

\bibitem{Affleck:1988wz}
I.~Affleck,
\newblock \emph{{Critical Behavior of SU($n$) Quantum Chains and Topological
  Nonlinear $\sigma$ Models}},
\newblock Nucl. Phys. B \textbf{305}, 582 (1988),
\newblock \doi{10.1016/0550-3213(88)90117-4}.

\bibitem{Lecheminant:2015iga}
P.~Lecheminant,
\newblock \emph{{Massless renormalization group flow in SU(N)$_k$ perturbed
  conformal field theory}},
\newblock Nucl. Phys. B \textbf{901}, 510 (2015),
\newblock \doi{10.1016/j.nuclphysb.2015.11.004},
\newblock \eprint{1509.01680}.

\bibitem{Yao:2018kel}
Y.~Yao, C.-T. Hsieh and M.~Oshikawa,
\newblock \emph{{Anomaly matching and symmetry-protected critical phases in
  $SU(N)$ spin systems in 1+1 dimensions}},
\newblock Phys. Rev. Lett. \textbf{123}(18), 180201 (2019),
\newblock \doi{10.1103/PhysRevLett.123.180201},
\newblock \eprint{1805.06885}.

\bibitem{Wamer:2019oge}
K.~Wamer, M.~Lajk\'o, F.~Mila and I.~Affleck,
\newblock \emph{{Generalization of the Haldane conjecture to SU($n$) chains}},
\newblock Nucl. Phys. B \textbf{952}, 114932 (2020),
\newblock \doi{10.1016/j.nuclphysb.2020.114932},
\newblock \eprint{1910.08196}.

\bibitem{Herviou:2023unm}
L.~Herviou, S.~Capponi and P.~Lecheminant,
\newblock \emph{{Even-odd effects in the $J_{1}-J_{2}$ SU(N) Heisenberg spin
  chain}},
\newblock Phys. Rev. B \textbf{107}(20), 205135 (2023),
\newblock \doi{10.1103/PhysRevB.107.205135},
\newblock \eprint{2302.14090}.

\bibitem{Lecheminant:2024apm}
P.~Lecheminant and K.~Totsuka,
\newblock \emph{{Lieb-Schultz-Mattis constraints for the insulating phases of
  the one-dimensional SU($N$) Kondo lattice model}}  (2024),
\newblock \eprint{2404.11030}.

\bibitem{Levin_2012}
M.~Levin and Z.-C. Gu,
\newblock \emph{Braiding statistics approach to symmetry-protected topological
  phases},
\newblock Physical Review B \textbf{86}(11) (2012),
\newblock \doi{10.1103/physrevb.86.115109}.

\bibitem{Li:2023mmw}
L.~Li, M.~Oshikawa and Y.~Zheng,
\newblock \emph{{Non-Invertible Duality Transformation Between SPT and SSB
  Phases}}  (2023),
\newblock \eprint{2301.07899}.

\bibitem{Li:2023knf}
L.~Li, M.~Oshikawa and Y.~Zheng,
\newblock \emph{{Intrinsically/Purely Gapless-SPT from Non-Invertible Duality
  Transformations}}  (2023),
\newblock \eprint{2307.04788}.

\bibitem{Chatterjee:2024ych}
A.~Chatterjee, O.~M. Aksoy and X.-G. Wen,
\newblock \emph{{Quantum Phases and Transitions in Spin Chains with
  Non-Invertible Symmetries}}  (2024),
\newblock \eprint{2405.05331}.

\bibitem{Chiu:2015mfr}
C.-K. Chiu, J.~C.~Y. Teo, A.~P. Schnyder and S.~Ryu,
\newblock \emph{{Classification of topological quantum matter with
  symmetries}},
\newblock Rev. Mod. Phys. \textbf{88}(3), 035005 (2016),
\newblock \doi{10.1103/RevModPhys.88.035005},
\newblock \eprint{1505.03535}.

\bibitem{Hagendorf_2012}
C.~Hagendorf and P.~Fendley,
\newblock \emph{The eight-vertex model and lattice supersymmetry},
\newblock Journal of Statistical Physics \textbf{146}(6), 1122–1155 (2012),
\newblock \doi{10.1007/s10955-012-0430-0}.

\bibitem{Hagendorf:2012fz}
C.~Hagendorf,
\newblock \emph{{Spin chains with dynamical lattice supersymmetry}},
\newblock J. Stat. Phys. \textbf{150}, 609 (2013),
\newblock \doi{10.1007/s10955-013-0709-9},
\newblock \eprint{1207.0357}.

\bibitem{Matsui:2016oqq}
C.~Matsui,
\newblock \emph{{Spinon excitations in the spin-1 XXZ chain and hidden
  supersymmetry}},
\newblock Nucl. Phys. B \textbf{913}, 15 (2016),
\newblock \doi{10.1016/j.nuclphysb.2016.09.002},
\newblock \eprint{1607.04317}.

\bibitem{Watanabe:2021wwt}
H.~Watanabe, M.~Cheng and Y.~Fuji,
\newblock \emph{{Ground state degeneracy on torus in a family of $\mathbb{Z}_N$
  toric code}},
\newblock J. Math. Phys. \textbf{64}(5), 051901 (2023),
\newblock \doi{10.1063/5.0134010},
\newblock \eprint{2211.00299}.

\bibitem{Hu_2023}
Y.~Hu and H.~Watanabe,
\newblock \emph{Spontaneous symmetry breaking without ground state degeneracy
  in generalized $n$-state clock model},
\newblock Physical Review B \textbf{107}(19) (2023),
\newblock \doi{10.1103/physrevb.107.195139}.

\end{thebibliography}


\end{document}